\documentclass[fleqn,10pt]{wlscirep}
\usepackage[utf8]{inputenc}
\usepackage[T1]{fontenc}
\usepackage{graphicx}

\usepackage{array}

\usepackage{bm}
\usepackage{textgreek}
\usepackage{comment}
\usepackage{siunitx}
\usepackage{amsmath}
\usepackage{color}
\usepackage{xr}
\usepackage{hyperref}
\makeatletter
\usepackage[font=small,labelfont=bf]{caption}
\title{Magnetic vortex writing and local reversal seeding in artificial spin-vortex ice via all-optical and surface-probe control}
\author[1,*,$\dagger$]{Holly Holder}
\author[1,2,3,*,$\dagger$]{Jack C. Gartside}
\author[1]{Alex Vanstone}
\author[4,5]{Troy Dion}
\author[1]{Xiaofei Xiao}
\author[1,2]{Kilian D. Stenning}
\author[1]{Tingjun Zheng}
\author[1]{Daniel Bromley}
\author[1]{Tobias Farchy}
\author[1]{Rupert F. Oulton}
\author[1,2]{Will R. Branford}

\affil[1]{Blackett Laboratory, Imperial College London, London SW7 2AZ, United Kingdom}
\affil[2]{London Centre for Nanotechnology, Imperial College London, London SW7 2AZ, United Kingdom}
\affil[3]{Institute for Materials Research, Tohoku University, Japan}
\affil[4]{Center for Science and Innovation in Spintronics (CSIS), Tohoku University}
\affil[5]{Solid State Physics Laboratory, Kyushu University, Japan}
\affil[*]{Corresponding author e-mails: h.holder20@imperial.ac.uk, j.carter-gartside13@imperial.ac.uk}
\affil[$\dagger$]{These authors contributed equally.}

\date{\today}
\begin{abstract}
Artificial spin-vortex ice (`ASVI') is a reconfigurable nanomagnetic metamaterial consisting of magnetic nanoislands tailored to support both Ising macrospin and vortex textures. ASVI has recently shown functional applications including reconfigurable magnonics and neuromorphic computing, where the introduction of vortex textures broadens functionality beyond conventional artificial spin ice which generally supports macrospin states. However, local control of writing vortex states in ASVI remains an open challenge. Here we demonstrate techniques for field-free magnetic vortex writing in ASVI. We expand ASVI to support metastable macrospin, single-vortex and double-vortex states. All-optical writing via focused laser illumination can locally write double-vortex textures, and surface-probe writing using an MFM tip can locally write single-vortex states. We leverage this writing to tailor and explore the reconfigurable energy landscape of ASVI, demonstrating programmable local seeding of avalanche-like reversal events. The global field-free texture-selective writing techniques reported here expand the suite of nanomagnetic control techniques, with a host of future applications including fundamental studies of avalanche dynamics, physical memory, and direct writing of nanomagnetic `weights' in physical neuromorphic neural networks.
\end{abstract}

\begin{document}
\maketitle

\section{Introduction}
Artificial spin ices (ASIs), so named due to the ice-rule framework used to investigate their vertex energies\cite{harris1997}, are systems of closely packed magnetic nanoislands whose competing geometrically-frustrated dipolar interactions result in a large number of low-energy microstates\cite{wang2006}. The broad design space of nanoisland shape/size and array geometry available in ASI nano-fabrication from 2D\cite{skaervo2020,schiffer2021} to 2.5D\cite{dion2024dipolar} and 3D systems\cite{berchialla2024focus} allows for a highly-tunable system, providing a window through which to study fundamental physics, long-range coupling, and emergent many-body behaviours\cite{skaervo2020,schiffer2021,ladak2010,morgan2011,bingham2021,chern2014,branford,sloetjes2024,zhang,zhang2013,jensen2024,gartside2018,saccone2023-2}. ASIs also provide a platform to study and realise functional architectures, such as reconfigurable magnonic systems\cite{kaffash,gliga2020dynamics,iacocca2020tailoring,bhat2020magnon,dion2019tunable,gartside2021,dion2022observation,dion2024dipolar,de2023tuning,bhat2025magnon}, and probabilistic and neuromorphic computing schemes\cite{arava,jensen,hon2021,gartside2022,stenning2023neuromorphic,hu,liang2024ultrafast}. 

Stadia-shaped magnetic nanoislands are traditionally designed with lengths and aspect ratios \cite{seynaeve} to preferentially support Ising macrospin textures\cite{wang2006}, in which the nanoisland magnetisation is oriented by shape anisotropy to lie parallel along the nanoisland long-axis. Magnetic vortex textures, topological defects typically observed in circular and elliptical islands, consist of a magnetic flux-closure pattern centred around a point-defect core (one core in the case of single-vortices, two cores for double-vortices)\cite{shinjo2000,metlov2002,gartside2022,guslienko2008,buchanan2005,zhang2010,saccone2023,tchernyshyov,vavassori2004,jain2010,sasaki2022,stenning2021,talapatra2020}. Vortex textures have far less stray field than macrospins due to their internal flux closure, and generally exhibit distinct switching fields compared with macrospin textures due to their differing magnetic structure and energetics. Previous studies have explored the local induction of topological magnetic defects via a scanning magnetic MFM tip\cite{gartside2016,gartside2018,stenning2021,romming2013writing,magiera2012magnetic,magiera2014magnetic}, and optical manipulation of magnetic topological defects in thin films \cite{finazzi,eggebrecht} and disks \cite{taguchi,fu}.

Recently a version of ASI was demonstrated in which nanoislands were geometrically tailored such that both macrospin and vortex textures are metastable -- so called `artificial spin-vortex ice' or ASVI\cite{gartside2022,stenning2023neuromorphic}. In ASVI macrospin and vortex states are engineered to be similar enough in energy such that both states can be accessed via uniform global field protocols, resulting in an enlarged microstate space\cite{gartside2022,stenning2023neuromorphic,saccone2023,dion2024dipolar}. The geometric tailoring is relatively simple, with the relative energies of macrospin and vortex textures determined by the nanoisland length-to-width aspect ratio and thickness\cite{gartside2022,saccone2023,dion2024dipolar}. The bi-stable textures expand the system beyond the Ising model\cite{louis2018,sklenar2019} and give rise to a host of emergent properties including enhanced physical memory, formation of separated domain-like regions of clustered vortex and macrospin nanoislands, and enriched magnon spectra. These properties, arising from introducing vortices alongside macrospins, have been leveraged for reconfigurable magnonics\cite{dion2024dipolar} and neuromorphic computing schemes\cite{gartside2022,stenning2023neuromorphic} including experiments demonstrating enhanced computational performance. 

A variety of interesting emergent dynamics and functional benefits arise from introducing metastable vortex states to ASI -- but a greater and more detailed range of experimental methodologies by which to locally write and control vortex states in nanoislands would benefit future devices. Macrospin states have a range of control protocols, including magnetic surface-probe tip-writing\cite{wang2016rewritable,gartside2018,lehmann2022poling}, all-optical switching\cite{stenning2023}, spin-wave control\cite{bhat2020magnon,baumgaertl2023reversal}, in addition to electrical/spin-torque switching approaches which can be challenging to achieve in dense strongly-coupled nanoisland arrays due to difficulties patterning many nanoscale electrical control lines. Vortex states lack the spatial symmetry breaking and net magnetisation of macrospin states, which makes devising control protocols harder, although different approaches have been devised \cite{gartside2016,gartside2018,stenning2021,romming2013writing,magiera2012magnetic,magiera2014magnetic,finazzi,eggebrecht,taguchi,fu}. In order to fully exploit the benefits of ASVI-style systems combining collinear macrospin and chiral vortex textures, protocols must be further developed which can locally and controllably write vortex states to nanoislands. Achieving this will open new possibilities in fundamental and applied studies, including seeding magnetic avalanches and direct programming of neuromorphic physical neural network weights.

Here we study an expanded ASVI hosting metastable macrospin, single-vortex and double-vortex states, and experimentally demonstrate two approaches to locally writing vortex textures in nanoislands: Single-vortex writing via scanning of a magnetic surface-probe MFM tip, and double-vortex writing via an all-optical approach using a focused laser spot, with selective control afforded by the laser polarisation. 

\begin{figure*}[t!]
\centering
\includegraphics[width=1.0\textwidth]{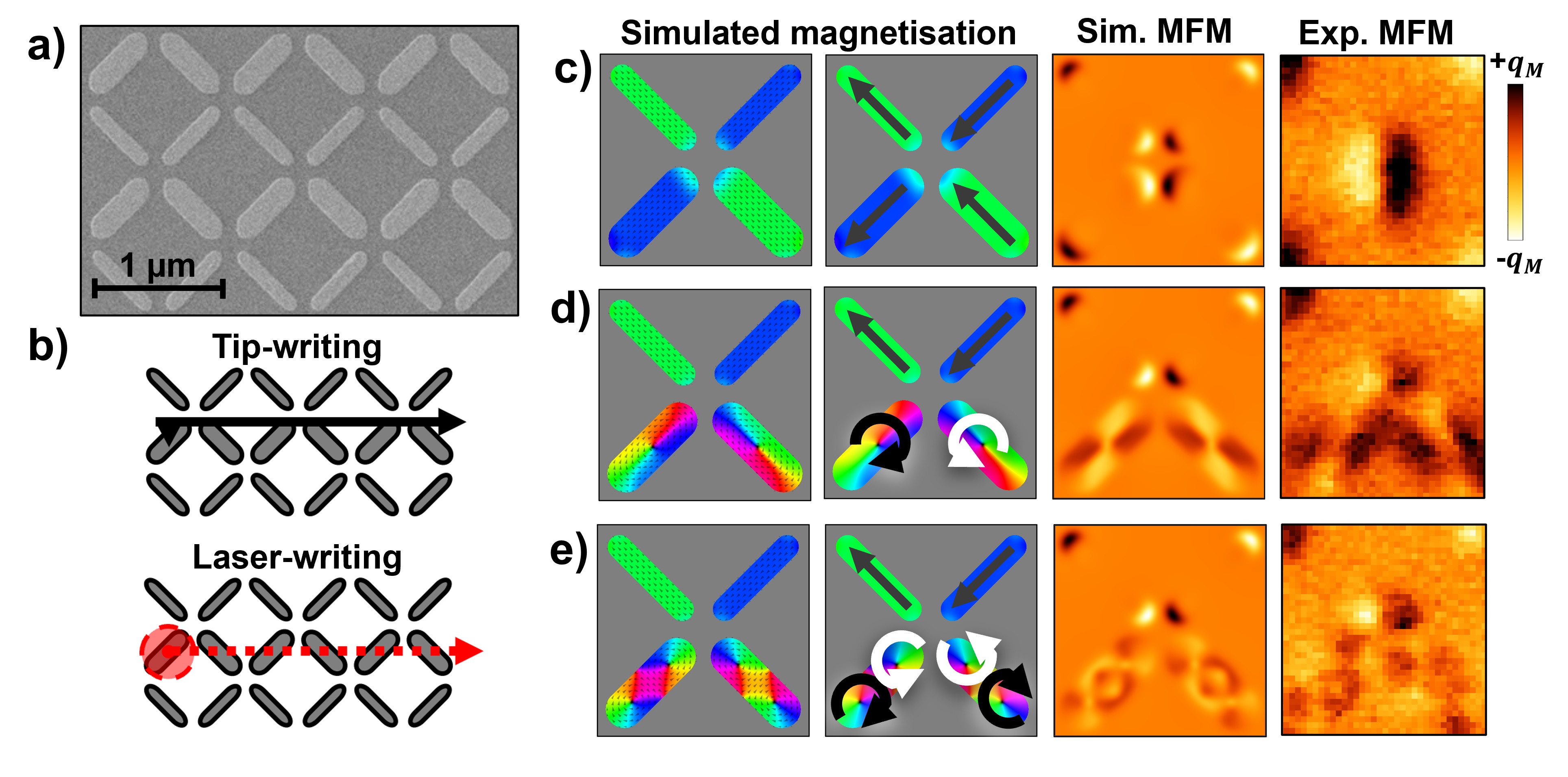}
\caption{\label{fig1} \textbf{Introduction to artificial spin-vortex ice hosting macrospin, single-vortex and double-vortex states.} \textbf{a)} SEM image of an ASVI Array. Array is `width-modified' with alternating rows of wide and narrow nanoislands. \textbf{b)} Schematic illustrating scanning probe (`Tip-writing'), and all-optical (`Laser-writing') technique protocols with respect to nanoisland geometries. \textbf{c-e)} Simulated micromagnetic textures, simulated MFM (60 nm lift height), and experimental MFM images of \textbf{c)} nanoisland macrospins, \textbf{d)} single-vortex states, and \textbf{e)} double-vortex states. All narrow nanoislands are in macrospin states. Arrows and colours represent the direction of magnetisation within each nanoisland. Simulated nanoislands have vertex gaps of 100 nm, lengths of 600 nm, and thicknesses of 20 nm, with widths of 200 nm, and 125 nm for wide and narrow nanoislands.}
\end{figure*}

\begin{table}
    \centering
    \begin{tabular}{|c|c|c|c|c|c|c|c|} \hline 
         [nm]& $\mathbf{L_{Wide}}$&  $\mathbf{W_{Wide}}$&  $\mathbf{R_{Wide}}$& $\mathbf{L_{Narrow}}$& $\mathbf{W_{Narrow}}$&  $\mathbf{R_{Narrow}}$& $\mathbf{V}$\\ \hline 
         \textbf{Array 1}& 546$\pm$11&  182$\pm$6&  3& 546$\pm$15& 122$\pm$11&  4.5&  91$\pm$9\\ \hline
         \textbf{Array 2}& 593$\pm$9&  203$\pm$6&  2.9& 588$\pm$10& 132$\pm$6&  4.5&  95$\pm$5\\ \hline
         \textbf{Array 3}& 569$\pm$12&  199$\pm$7&  2.9& 569$\pm$14& 135$\pm$11&  4.2&  86$\pm$9\\ \hline
         \textbf{Array 4}& 595$\pm$10&  213$\pm$4&  2.8& 596$\pm$11& 145$\pm$13&  4.1&  86$\pm$9\\ \hline
         \textbf{Array 5}& 600&  200&  3& 600& 125& 4.8&  100\\ \hline
    \end{tabular}
    \caption{Average lengths (`L'), widths (`W'), aspect ratios (`R'), and vertex gaps (`V', from a nanoisland end to a vertex centre) of wide and narrow nanoislands in the arrays used within this work, measured with a Scanning Electron Microscope. Array 5 is a large-area array. The values reported for Arrays 1-4 are reasonably representative of the full arrays due to their overall size, but the much larger size of Array 5 makes reporting representative errors difficult. The values stated for the Array 5 are therefore ideal, but fabrication-based quenched disorder will affect the behaviour of the array.}
    \label{dims}
\end{table}

\section{Results and Discussion}

\subsection*{Artificial spin-vortex ice arrays}

The ASVI arrays used here are `width-modified’, consisting of alternating rows of wide and narrow stadia nanoislands as can be seen in the SEM image in Figure \ref{fig1} a). Wide nanoislands are metastable and host single-vortex, double-vortex and macrospin states. Narrow nanoislands almost exclusively host macrospin states. Breaking the array into these two nanoisland types affords us additional global control of array microstates as narrow nanoislands have significantly higher switching fields, and allows the possibility of 1D avalanches to be investigated in the future \cite{bingham2021}. A range of arrays are investigated with differing nanoisland dimensions. Arrays 1-4 are 14 \textmu m $\times 14$ \textmu m and comprise 26 rows (13 wide and 13 narrow) with 26 nanoislands per row. A large area (mm$^{2}$) array was also studied, labelled as Array 5. The dimensions of each array are listed in Table \ref{dims} (see Supporting Information Figures S1-4 for geometric analysis). Nanoislands are 20 nm thick Permalloy (Ni$_{81}$Fe$_{19}$) with a 5 nm Al\textsubscript{2}O\textsubscript{3} cap.

Figure \ref{fig1} b) schematically illustrates the two vortex writing techniques used in this work; scanning probe (`Tip-writing') control via interaction between nanoislands and a Magnetic Force Microscopy (`MFM') Tip, and all-optical (`Laser-writing') via scanning of a continuous-wave, linearly-polarised focussed laser spot. Examples of wide nanoisland macrospin, single-vortex, and double-vortex textures are shown in Figure \ref{fig1} c-e) respectively, with MuMax3\cite{mumaxa} simulated magnetisation states, MuMax3 simulated MFM images and experimental MFM images shown in columns left-to-right.

\begin{figure*}[t!]
\centering
\includegraphics[width=1.0\textwidth]{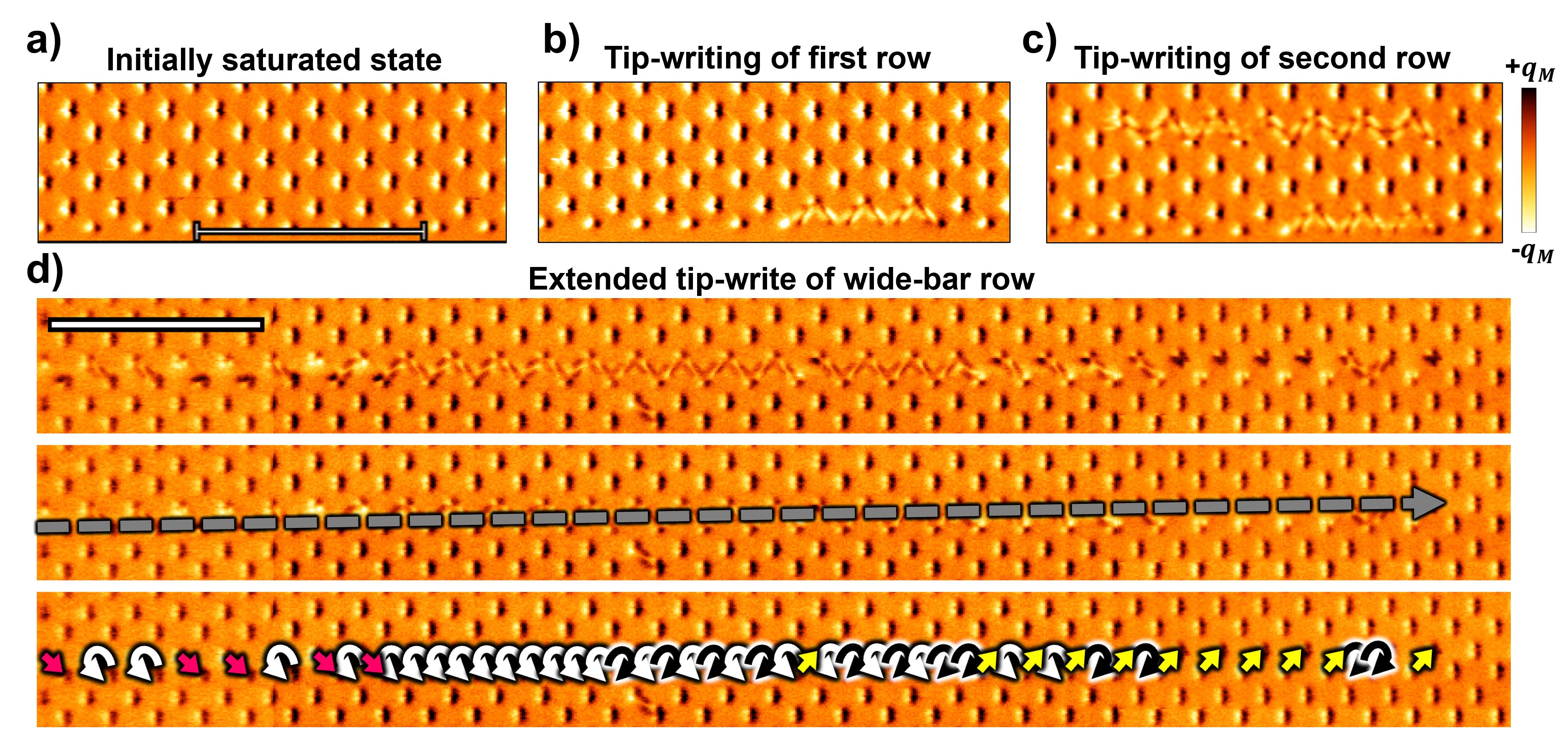}
\caption{\label{fig2} \textbf{Demonstration of tip-writing of single-vortex states.} \textbf{a)} MFM image showing section of Array 2 after initial global field saturation in the negative X-direction. Scale bar = 5 \textmu m. \textbf{b)} MFM image after tip-writing the bottom wide nanoisland row. A chain of single-vortices has been induced. \textbf{c)} MFM image after a second tip-write along the third wide nanoisland row. A long chain of primarily single-vortices has been induced, with a double-vortex in the middle of the chain and on the right end of the chain. \textbf{d)} MFM image showing section of Array 5 after tip-writing, with the approximate deduced tip motion path highlighted (middle row), and annotated vortex chiralities and macrospin-to-macrospin reversals (bottom row). A crossover in vortex writing behaviour is observed as the tip hits different nanoisland regions, going from purely anti-clockwise chirality on the left of the write-line to mixed chirality. Scale bar = 5 \textmu m.}
\end{figure*}

\subsection*{Surface-probe control of vortex textures}
We first demonstrate surface-probe writing of vortices with a magnetic tip. Here, a high-moment MFM tip is scanned in a line across wide nanoislands, primarily writing single-vortices. Figure \ref{fig2} a) shows a magnetic force micrograph of ASVI Array 2 in an initial global-field saturated state (negative X-direction), with Figure \ref{fig2} b) showing the same section after tip-writing along the bottom wide-bar row. A chain of six nanoislands has been written into single-vortex states by the MFM tip. Figure \ref{fig2} c) shows the magnetic state after a second tip-write along the third-from-bottom row; a chain of nanoislands written into primarily single-vortex states, with two double-vortices written at the middle and right side of the chain. Tip-writing of vortices with local spatial control is demonstrated, following previous work where the tightly-focused divergent monopole-like dipolar field at the apex of the high-moment MFM tip injects a vortex-core topological defect into the magnetic material, localised directly under the tip\cite{gartside2016,gartside2018,magiera2012magnetic,magiera2014magnetic} -- here the vortex core remains stable in the nanoisland after the MFM tip has moved away due to the ASVI aspect ratio engineering, which enables vortex texture stability.

Figure \ref{fig2} d) shows a section of Array 5 after tip-writing from left to right, in which the two lower images are duplicates of the top with added annotations. The middle row shows the approximate deduced trajectory of the tip movement path with a grey dashed line, and the bottom row shows the annotated written states of written vortices (anti-clockwise chirality in white, clockwise in black), along with the nanoisland orientations of tip-written macrospin states (-45$^{\circ}$ nanoislands with south-east macrospins in pink, +45$^{\circ}$ nanoislands with north-east macrospins in yellow). The tip trajectory is deduced to cross the lower end of the nanoisland at the left-hand-side of the scan line, with the point at which the tip crosses the nanoisland gradually moving upwards until it crosses the top end of the nanoisland at the right-hand-side of the scan line. A transition in the tip-written magnetisation states is observed across the scan line, going from writing anti-clockwise single-vortices (white) on the left, to a mix of anti-clockwise and clockwise single-vortices in the middle, to the possible start of a clockwise vortex (black) regime on the right. This suggests that the position at which the tip crosses the nanoisland has an impact on the chirality of the resultant written vortex state. Additionally, a change is observed in which nanoislands are written into oppositely-magnetised macrospin states, with macrospin reversals occurring only in nanoislands orientated at -45$^{\circ}$ to the horizontal (pink, with south-east pointing macrospins) on the left, and subsequent macrospin reversals on the right only in +45$^{\circ}$ nanoislands (yellow, with north-east pointing macrospins). Further examples showing evidence of the impact of the tip-nanoisland crossing point on the resultant written state, with some deviations from the trend, are shown in Supporting Figures S5 and S6.

These observations correspond with previous micromagnetic simulation work investigating the control over vortex chirality provided by tip-writing where a scanning magnetic tip crosses a nanomagnetic disk at different spatial positions\cite{stenning2021}. The success rate and fidelity of the tip-writing technique has room for improvement. Often attempted tip-writes give no writing (an `unsuccessful' write), and when writing does successfully occur, some islands along a write-line may not be written at all, or a reversed macrospin is written instead of a vortex (a `low fidelity' write). We have observed examples of writing events which inject vortices into many successive nanoislands, such as the written states shown in Figure \ref{fig2} b-c). It is speculated that the success and fidelity of tip-based vortex writing may depend on several parameters, such as the point of the nanoisland which the tip crosses, tip height above the nanoisland, scan angle, scan speed, and the quality of the tip itself (magnetic field amplitude and spatial conformation of the field), even between tips with the same nominal manufacturer-specified moment. As better control is gained over these parameters, we predict that fidelity and success will likely improve.

\begin{figure*}[t!]
\centering
\includegraphics[width=1.0\textwidth]{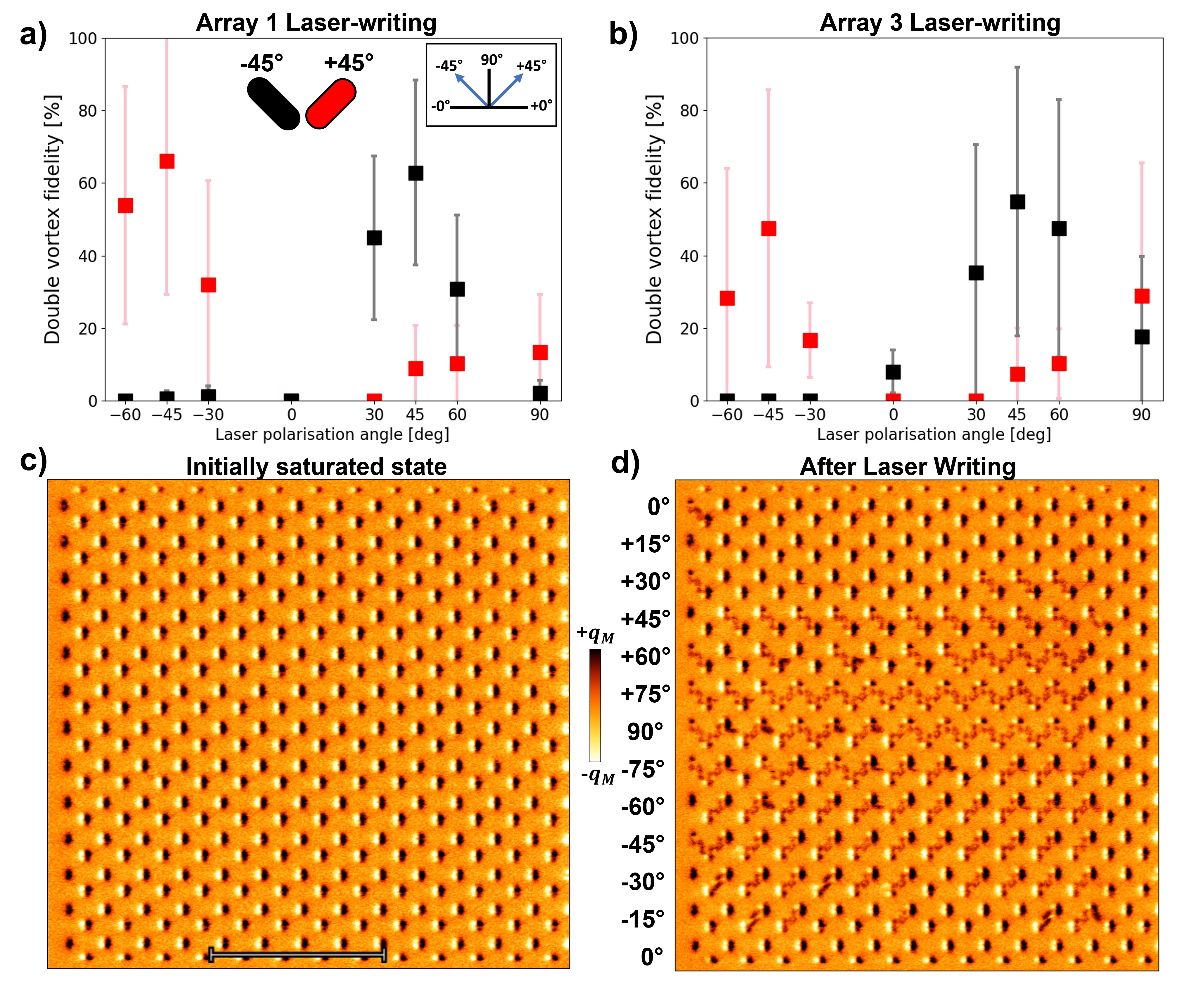}
\caption{\label{fig3} \textbf{Statistical investigation of laser-writing technique} \textbf{a-b)} Experimental fidelity of double-vortices induced by laser writing as a function of laser polarisation, for wide nanoislands from Arrays 1 and 3, with P = 4.8 mW. Statistics for nanoislands oriented at +45$^\circ$ are shown in red, while nanoislands oriented at -45$^\circ$ are shown in black. Error bars show the standard deviation. Insets show a schematic of two wide nanoislands showing orientation nomenclature and corresponding colours, and axes showing nomenclature of laser polarisation respectively. \textbf{c)} MFM image showing the initially saturated state of Array 4. Scale bar = 5 \textmu m. \textbf{d)} MFM image showing the same array after laser-writing of all wide nanoisland rows (P = 5.0 mW). The laser polarisation is labelled on the left hand side of each row.}
\end{figure*}

\subsection*{All-optical control of vortex textures}
We now consider a second method for local writing of magnetic vortices; all-optical switching. Here, a linearly-polarised continuous-wave focused laser spot is scanned across rows of wide nanoislands, primarily writing double-vortex states. The all-optical writing of double-vortices offers polarisation-based selectivity of which $\pm45^{\circ}$ nanoisland orientations are written.

Laser-writing statistics were collected for Arrays 1 and 3. Figures \ref{fig2} a) and b) show the fidelities of double-vortex writing in these two arrays, with inset schematic in a) illustrating the nomenclature for nanoisland orientation and laser polarisation. The percentage of wide nanoislands illuminated at laser power P = 4.8 mW that were laser-written into double-vortices is shown as a function of laser polarisation for both $\pm$45$^{\circ}$ angled nanoisland orientations, denoted by red ($+45^{\circ}$) and black ($-45^{\circ}$) squares. Only laser polarisations with at least 4 repeated illuminated rows are included, with the greatest number of illuminations performed at $\pm$45$^{\circ}$. The all-optical writing preferentially writes double-vortices into nanoislands where the laser polarisation is aligned closely to the short axis (width) of the nanoisland. Laser-writing can also induce nanoislands into single-vortex states, though these were observed more rarely (see Supporting Information Tables S1 and S2 for full statistics, including rare occurrences of damage).

The total double-vortex writing fidelity shows some variation between the two array geometries with Array 1 exhibiting higher fidelity, but both arrays show the same qualitative behaviour between fidelity and laser polarisation angle. Further investigations can determine if these fidelity discrepancies are caused by the differences in nanoisland dimensions between arrays 1 and 3, with the possibility of optimising nanoisland dimensions for enhanced fidelity. A clear trend in fidelity can be seen in Figures \ref{fig3} a) and b) for both arrays, where double-vortex formation is preferential when the laser polarisation is aligned along the nanoisland short axis (width). To demonstrate more granular variations of the laser polarisation on all-optical double-vortex writing, we perform a study where adjacent rows of an ASVI array (Array 4) are written with $15^{\circ}$ steps of polarisation between rows. Figure \ref{fig3} c) shows an MFM image of Array 4 in an initial global-field saturated state (negative X-direction). Figure \ref{fig3} d) shows the same array after laser-writing all wide-bar rows, with the laser polarisation shown at the left-hand side of each written row. The effect of the laser polarisation on the all-optical double-vortex writing can be clearly observed -- $\pm$45$^{\circ}$ polarisation gives high-fidelity writing in the angled nanoisland subset where the nanoisland width is parallel to the laser polarisation, $75-90^{\circ}$ shows good fidelity writing in both nanoisland subsets, while $0^{\circ}$ polarisation results in almost no writing events. It is noted that the illumination shown in Figure \ref{fig3} c-d) was not scanned to the very right-hand-side of the array, explaining the lack of events in that region, and the resulting longer exposure of the final illuminated column resulting in two damaged nanoislands (See Supporting Information Figure S7 for re-saturated image).

The strong polarisation-dependence of which nanoisland orientation is written may be due to polarisation-dependent optical absorption within the nanoislands. To test this, simulations using the Lumerical FDTD package \cite{lumerical} were performed to calculate the total optical absorption within the nanoislands. The absorption was found to be uniformly higher when the laser polarisation was oriented along the short axis (width) of the nanoisland, across Arrays 1 - 3, although the increase in absorption was relatively small. Array 1 wide nanoisland absorptions were calculated to be 5.4$\%$ (perpendicular) and 4.8$\%$ (parallel), and 9.1$\%$ and 7.8$\%$ for Array 3 (see Supporting Information Figure S8). However, it is unclear why there is a difference in fidelity between $0^{\circ}$ and $90^{\circ}$ polarisations, which are symmetric from an optical absorption/polarisation standpoint and only have their degeneracy lifted when considering the magnetic state. Further investigation of the differences in the all-optical writing behaviour between these polarisations provide fertile ground for future work, including studying well-separated nanoislands which are not influenced by the magnetic state/stray field of neighbouring bars.

The differences in optical absorption at different polarisations appear to be relatively small, and although geometrical imperfections will alter experimental absorption it is speculated that the polarisation selectivity may also be influenced by an additional physical mechanism, such as a magneto-optical effect. A single optically induced macrospin-to-macrospin reversal was also observed. The physical mechanism(s) underpinning the optical writing of double-vortices is an area of ongoing research.

Bars showing signs of magnetic damage after illuminations (determined by reduced MFM image contrast) were not included in these switching statistics, nor were nanoislands that were induced into single-vortex/indeterminate multi-domain textures.

\subsection*{Local seeding of avalanche-like reversal via double-vortex writing}

\begin{figure*}[t!]
\centering
\includegraphics[width=1.0\textwidth]{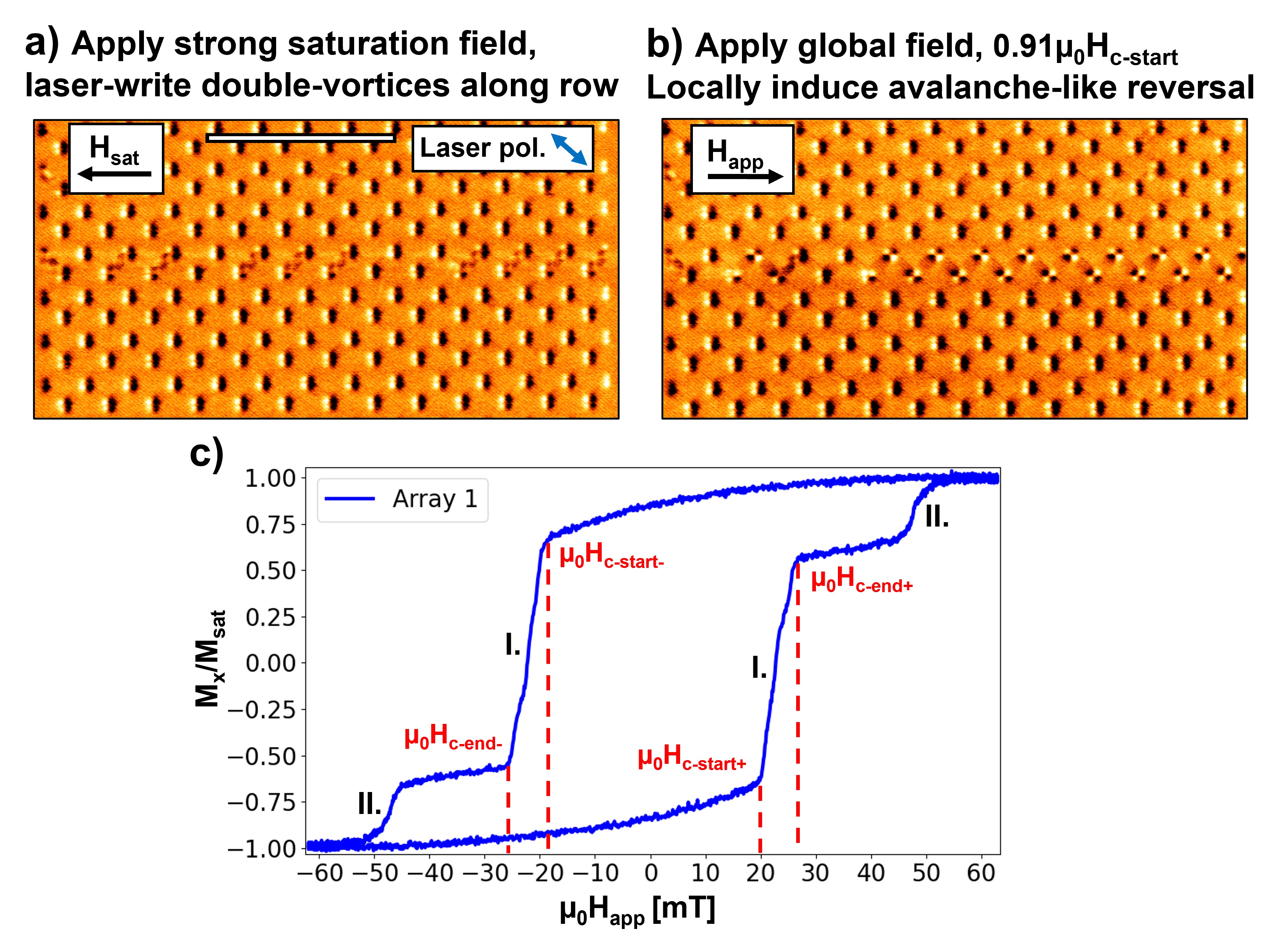}
\caption{\label{fig4} \textbf{Locally-seeded avalanche-like reversal via optical double-vortex writing.} \textbf{a)} Experimental MFM image showing nine laser-written double-vortices along a row in Array 1, following application of an initial global saturating field (negative-X direction). Scale bar is 5 \textmu m, laser polarisation shown inset. \textbf{b)} Subsequent field-activated microstate evolution of same array section. A positive X-direction global magnetic field is applied at 91$\%$ of \textmu$_0$H\textsubscript{c-start}. Around the double-vortices local modification of the stray field landscape leads to an avalanche-like chain of reversals occurs leading to a one-dimensional row of type-1 ASI vertices. Two isolated switches occurred in regions of the array away from double-vortices. \textbf{c)} Magnetic hysteresis loop for Array 1 with start and end switching fields labelled for positive and negative magnetic fields. Regions `I.' are the switching (positive and negative) of wide nanoislands, with regions `II.' the switching of narrow nanoislands.}
\end{figure*}

We now examine the ability of optically-written double-vortices to influence the behaviour of neighbouring nanoislands in the array, controllably seeding switching events of unwritten macrospin islands in a manner similar to one-dimensional avalanche-like reversal. Double-vortex states have very low stray field and a significantly lower coercive switching field than macrospin states, allowing them to locally modify the stray field landscape and hence switching dynamics of the array, in addition to their own lowered coercive field. The lower coercive field of the double-vortices compared to macrospins is the opposite effect to single-vortices which have higher coercive fields relative to macrospins, as previously demonstrated\cite{gartside2022}. We first leverage this behaviour to demonstrate how optically-written double-vortices can be used to influence the switching behaviour of neighbouring nanoislands and controllably seed avalanche-like reversal events which propagate through neighbouring unwritten macrospin states, and then discuss the details behind the coercive field and local stray field modification in the subsequent section.

We begin by laser writing double-vortex states along a single row of otherwise global-field saturated ASVI (Array 1, saturated in negative X-direction). Figure \ref{fig4} a) shows the laser-written state, written at a laser power of 4.8 mW and polarisation +45$^\circ$. Figure \ref{fig4} b) then shows the same section of the array after the application of a positive +X-direction field, shown by black arrow) global magnetic field at 18.2 mT, 91$\%$ of \textmu$_0$H\textsubscript{c-start} = 20 mT where H\textsubscript{c-start} is the field at which macrospins begin to switch as indicated in the Magneto-Optical Kerr Effect measured M(H) loop in Figure \ref{fig4} c). This applied field of 18.2 mT, below the macrospin switching field, is chosen such that switching events will only occur where the presence of optically-written double-vortices have locally modified the array switching dynamics. The resultant state after applying 18.2 mT, seen in Figure \ref{fig4} b), shows two interesting effects: optically-written double-vortices have switched to +X-direction macrospins, due to the low coercive field of double-vortex states relative to macrospin states, and previously unwritten macrospins adjacent to double-vortex nanoislands have also reversed their state. This leads to the formation of a one-dimensional domain of type-1 ASI vertices, the ASI ground state which can be challenging to prepare\cite{wang2006,gartside2021,morgan2011}, along the path of the laser-written line. N.B. One wide nanoisland, possibly showing signs of slight damage, and one wide nanoisland at the array edge -- possibly due to the absence of stray field from one side (not shown -- see Supporting Information Figure S9) also switched under application of the 18.2 mT field, and a single-vortex was also induced. No other of the 338 wide nanoislands in the array switched under application of 18.2 mT besides those seen in the image, highlighting the local nature of the switching control. Fidelity of the double-vortices switching back to macrospins under application of 18.2 mT field is not perfect, with two double-vortices remaining at the left-hand side of Figure \ref{fig4} b), likely due to nanofabrication imperfections leading to a distribution of switching fields between different islands, or possible slight damage.

This result is intriguing, as not just the optically-written double-vortex islands were affected by the applied global field, but also adjacent unwritten macrospin states which have had their reversal dynamics modified by local changes to the stray field landscape caused by the presence of double-vortices. A chain of five adjacent unwritten macrospin states (with a double-vortex at each end of the chain) were all switched by the 18.2 mT field, and the resultant type-1 domain chain is 17 nanoislands long with only 7 nanoislands originally written into a double-vortex state. This propagating switching behaviour may be considered a type of one-dimensional avalanche reversal\cite{bingham2021} and the optically-written double-vortices may be considered to act as locally-seeded defects which trigger the propagating reversal event. To the best of our knowledge, this is the first demonstration of such reconfigurable locally-seeded switching where manually written nanoisland states exert control over longer-range switching behaviours.

In repeated tests (see Supporting Figures S10-12) the same effect was observed, in which double-vortices were laser-written in different locations within the array, and were switched into macrospin states at fields below \textmu$_0$H\textsubscript{c-start}, also triggering macrospin-to-macrospin reversals of neighbouring bars, although further large-scale avalanches were not observed. The different locations of the optically-written double-vortices throughout the array on subsequent writes confirms that this effect is not solely due to particular nanoislands having a lower switching field. In these repeat experiments the same aforementioned edge nanoisland also reversed, and the other noted nanoisland induced into a single-vortex, suggesting that these nanoislands are specifically anomalous.

Figure \ref{fig4} c) shows the corresponding hysteresis loops of Array 1, with the start and end values of the switching fields annotated. Regions `I.' label the reversal slope of wide nanoislands, while Regions `II.' the narrow nanoislands. Hysteresis loops are calibrated and scaled to known switching field values.

\subsection*{Multi-level switching fields of macrospin, single-vortex and double-vortex states}

\begin{figure*}
\centering
\includegraphics[width=1.0\textwidth]{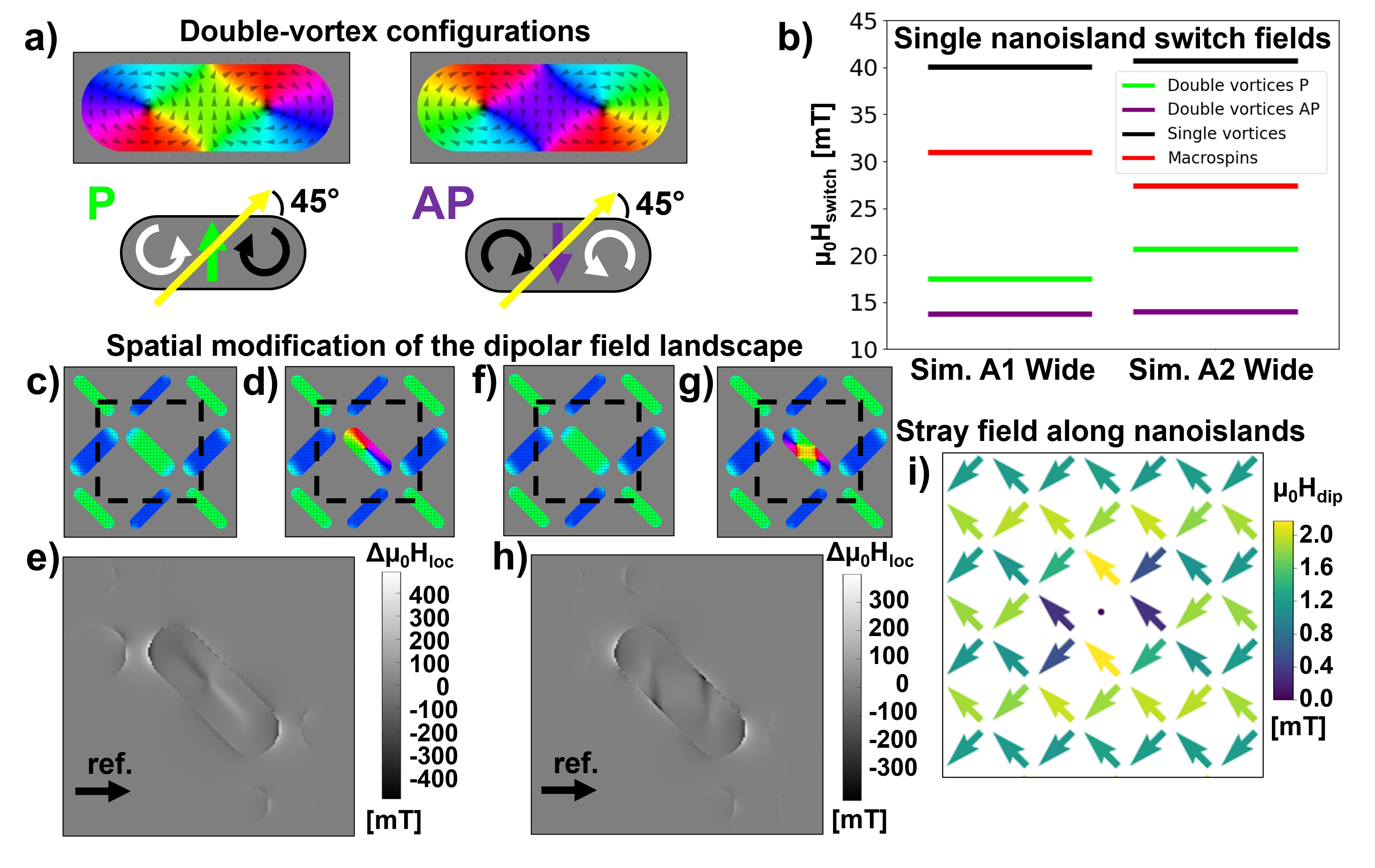}
\caption{\label{fig5} \textbf{Simulated switching fields of different nanoisland textures and modification of the stray field landscape} \textbf{a)} Micromagnetic simulations (above) and corresponding schematics (below) showing texture of the two double-vortex nanoisland configurations used to calculate switching fields. Configurations are labelled as `P' (parallel) or `AP' (anti-parallel) depending on direction of the central component of the double-vortex texture with respect to the applied magnetic field at +46$^{\circ}$ (yellow arrow). N.B. simulated fields are applied at $\pm$45$^{\circ}$ (to break symmetry) and then averaged, corresponding to the experimental sample geometry in which nanoislands lie at $\pm$45$^{\circ}$ and fields are applied parallel to the X-axis. \textbf{b)} Plot showing the simulated magnetic switching field of a single-vortex (black), parallel double-vortex (lime), anti-parallel double-vortex (purple), and macrospin state (red) of a single wide nanoisland, for dimensions corresponding to `wide' nanoislands from  Array 1 (left column) and Array 2 (right column). \textbf{c-d)} Cropped array section of micromagnetic simulations in an all-macrospin state, and with a clockwise single-vortex, corresponding to dimensions of Array 5. \textbf{e)} Plot of matching array section to that shown by dotted boxes in c-d), with greyscale intensity representing the difference in local demagnetising field along the X-direction between the two states. \textbf{f-g)} Cropped array section of micromagnetic simulations in an all-macrospin state, and with a double-vortex, corresponding to dimensions of Array 5. \textbf{h)} Plot of matching array section to that shown by dotted boxes in f-g), with greyscale intensity representing the difference in local demagnetising field along the X-direction direction between the two states. \textbf{i)} Plot showing section of ASVI Array 5 model used to calculate reduction in stray magnetic field in nanoislands neighbouring a vortex texture (shown here as a small dot). \textmu$_0$H\textsubscript{dip} is the magnitude of the stray field along the direction of magnetisation of each bar.}
\end{figure*}

The switching fields for different initial nanoisland magnetic textures (double-vortex, single-vortex, macrospin) into a macrospin aligned with the applied field were calculated via micromagnetic simulation. Note that simulations correspond to the experimental setup of the system, i.e. with nanoislands at $\pm$45$^{\circ}$ and applied fields along the X-direction (all fields had 1$^{\circ}$ symmetry breaking in the X-Y plane). The two double-vortex configurations can be seen in Figure \ref{fig5} a), with accompanying magnetic schematics. One is labelled `P’ due to the central region of the double-vortex texture containing a component aligned parallel to the applied magnetic field applied along +45$^{\circ}$. The other configuration is therefore `AP’ (anti-parallel) for the same reason.

For laser writes performed on horizontally saturated arrays the configuration of the written double-vortex states depended only on the initial saturation direction of the array, regardless of laser polarisation or laser scanning direction. The resulting configurations were such that laser writing of an array induced double-vortices with central components opposite to the saturation direction. Preliminary observations of an initially randomised array microstate suggests that the surrounding local stray field landscape may influence the resulting double-vortex configuration. A number of different state changes were induced by laser writing, including single-vortex-to-double-vortex and single-vortex-to-macrospin transformations, in addition to the previously mentioned macrospin-to-double-vortex and macrospin-to-single-vortex seeding (see Supporting Information Figure S13) 

Figure \ref{fig5} b) shows the resulting simulated switching fields of isolated wide nanoislands in initially macrospin, single-vortex, parallel double-vortex, and anti-parallel double-vortex states, for dimensions corresponding to Arrays 1 and 2. Further details can be seen in Supporting Figures S14-16. single-vortex textures are found to uniformly require much higher switching fields, as expected from previous work\cite{gartside2022}, with both clockwise and anti-clockwise chiralities showing identical switching fields. In contrast, simulated parallel and anti-parallel double-vortices are calculated to have quite different switching fields. Narrow nanoislands are shown computationally and experimentally to have a higher onset of macrospin-to-macrospin switching fields than even single-vortex wide nanoislands\cite{gartside2022}. The effect of geometry changes on the resulting switching fields can be seen, though the qualitative behaviour remains consistent between the two arrays.

Figure \ref{fig5} c) shows a cropped section of a simulated Array 2 (with narrow nanoislands simulated to be the same length as the wide bars), with all wide nanoislands in macrospin states, and d) the same state with the addition of a clockwise single-vortex. Simulations were performed with further columns to both the left and right-hand sides, all in macrospin states, to negate any edge effects. Figure \ref{fig5} e) shows a plot of the local difference in the demagnetising field (along the X-axis) between the vortex state, and the all-macrospin state, of the region highlighted by dotted boxes in c-d), where \textDelta \textmu$_0$H\textsubscript{loc} $>$ 0 represents a reduction in demagnetising field for the vortex state in the initial saturation direction (-X). Further information, including Y field plots, can be found in Supporting Figures S17-18. Similarly, Figure \ref{fig5} f-g) show the macrospin and double-vortex states, with Figure \ref{fig5} h) showing the difference in demagnetising field between the two states as before. The presence of a double-vortex can be seen to have a similar effect to that of a single-vortex on neighbouring bars. 

Vortex states exhibit internal flux-closure, and hence emit far less stray magnetic field than macrospin textures. This leads to modification of the effective switching fields of nanoislands which neighbour vortex textures, since the effective switching field \textmu$_0$H\textsubscript{c-eff} is a combination of the intrinsic nanoisland switching field \textmu$_0$H\textsubscript{c-int} (the field it would reverse at if it was an isolated nanoisland) and the effect of the stray dipolar field emanating from neighbouring nanoislands \textmu$_0$H\textsubscript{dip}, giving \textmu$_0$H\textsubscript{c-eff} = \textmu$_0$H\textsubscript{c-int} + \textmu$_0$H\textsubscript{dip}. 

In a saturated macrospin state where all nanoislands have been polarised by a global magnetic field (the so-called `type-2' artificial spin ice state), the stray field of neighbouring nanoislands increases the effective switching field \textmu$_0$H\textsubscript{c-eff}, as an applied field oriented against the nanoisland magnetisation would have to overcome both the intrinsic nanoisland switching field \textmu$_0$H\textsubscript{c-int} and the additional stray dipolar field from neighbouring magnets \textmu$_0$H\textsubscript{dip}. 

Figure \ref{fig5} i) shows a plot of the simulated stray dipolar field projected along the easy-axis (length) in the magnetisation direction, with nanoislands of dimensions matching Array 5. The central island is in a vortex state (represented by a dot), with zero stray field assumed.

Comparing \textmu$_0$H\textsubscript{c-eff} of a wide nanoisland next to the vortex to \textmu$_0$H\textsubscript{c-eff} of a nanoisland in a saturated type-2 state, we see a reduction in \textmu$_0$H\textsubscript{c-eff} of 1.6 mT, calculated by considering the projection of local dipolar fields along the nanoisland easy/long axis. 

As shown in Figure \ref{fig5} i), \textmu$_0$H\textsubscript{dip} varies depending on a nanoislands distance from the vortex site. N.B., previously we have considered applied fields along the `x-axis', at 45 degrees to all nanoislands. Here, we quote the dipolar field and effective field reduction along the nanoisland easy/long axis.

These modifications to the dipolar field landscape enable vortex textures to influence reversal pathways throughout the array\cite{leo,gartside2022}, enabling the local control of array switching by written vortices as shown in Figure \ref{fig4}.

The experimental switching behaviour of different nanoisland textures generally agrees with these simulated switching fields, highlighting the lower switching field of the parallel double-vortices relative to macrospin islands in a saturated type-2 state, as shown in Figure \ref{fig4} b). The exact switching fields of nanoislands in an array depend on the surrounding stray dipolar field landscape, the geometry of the bar, and experimental fabrication imperfections in nanoisland size, shape and material quality (often termed `quenched disorder').

Artificial spin ice has been used as a platform in which to study avalanche-like reversal dynamics\cite{drisko2017,chern2014,mengotti,ladak2010,bingham2021}. Although single and double-vortices have a similar effect on the local dipolar field landscape due to their internal flux-closure, reducing the effective switching field of neighbouring bars, their behaviour under applied magnetic field is very different. Double-vortex states have a significantly lower switching field than macrospins (Figure \ref{fig5} b) allowing them to serve as a `sacrificial template' where they can locally modify array switching behaviour and seed reversal events as shown in Figure \ref{fig4} a-c), switching to macrospin states in the process and leaving no vortex states behind. Conversely, single-vortex states have a significantly higher switching field than macrospins (Figure \ref{fig5} b), so in an operating window of global field amplitude above the macrospin switching field and below the single-vortex destruction field, can act as a permanent or reusable template in modifying field-driven switching processes of macrospins. 

%The local writing of different vortex textures presented in this study opens new, rich avenues into avalanche-like reversal studies, enabling direct on-demand seeding of point-defects in the magnetic landscape to controllably trigger avalanche-like dynamics.

\section{Conclusions}
In this work we have presented two methods for reconfigurable local control of single and double-vortex states in nanomagnetic, and expanded the range of nanoisland textures in artificial spin-vortex ice to include double-vortices, in addition to macrospin and single-vortex states, granting six states per nanoisland and a large $6^N$ microstate space in an $N$-island array (core polarity is not considered here). We have characterised the fidelity and polarisation-dependent selectivity of all-optical double-vortex writing, and shown evidence that surface-probe writing can offer chirality and macrospin/single-vortex writing selectivity via the point at which the magnetic tip crosses the nanoisland. 

The expanded degree of local microstate control demonstrated here enhances the possibilities for fundamental and applied studies harnessing the expanded range of physical dynamics in multi-texture magnetic systems such as ASVI. Our observation of locally-seeded propagating reversal events presents exciting opportunities to explore rich avalanche reversal dynamics in ASVI and related strongly coupled systems, with magnetic vortex textures acting as programmable defects in the stray dipolar field landscape which can trigger long-range avalanche-like reversal events. 

The possibilities here are considerable, prior to this work the study of avalanche-like dynamics in artificial spin systems relied on observation of uncontrolled stochastic nucleation of switching, which leads to challenging experimental considerations -- imaging of magnetic states is slow, and the many repeated imaging experiments required for stochastic processes can render experimental studies untenable. Here, our local writing of vortex textures provide reconfigurable seeding of reversal events, opening rich new avenues for studies into avalanche-like switching and related emergent many-body dynamics. While our methodology still has a degree of stochasticity, the spatial writing demonstrated here offers vastly improved experimental control relative to prior global-field based studies which lack local writing.

The customisation of the local dipolar field landscape and array switching pathways offers routes forward for applied studies, including data input of future ASVI-based neuromorphic and probabilistic computation schemes. Gaining local, reconfigurable control over the nucleation of magnetic vortex textures in nanomagnetic arrays enables direct exploitation of the enriched microstate dynamics and avalanche physics on offer in strongly-interacting nanomagnetic arrays. We anticipate future work using the methods described here for programming computational states, both local writing of nanoislands to implement physical neuromorphic network weights, and preparation of switching pathways to implement Ising/Boltzmann machine-like optimisation computing and probabilistic computing schemes.

\section{Methods}
\subsection*{Fabrication}
ASVI arrays were fabricated via an electron beam lithography (Raith eLine) lift-off process using PMMA resist. Sample measurements were determined via Scanning Electron Microscopy, on the same Raith system, with average values reported for Arrays 1-3, where stated uncertainties are the standard deviation from this average in a sampled nanoisland distribution.
\subsection*{Laser-writing}
Laser-writing was performed with a \textlambda\space = 633 nm continuous-wave (CW) HeNe laser via a Witec alpha 300R microscopy system, focused through a 100x (0.9 N.A.) Zeiss objective, with a final Gaussian spot $\frac{1}{e^2}$ diameter of 580 nm. Transmission through the lens is stated to be 91$\%$. As such, powers reported in this work have been multiplied by 0.91 after power measurement without the objective lens present (some further optical losses may be present in the system). A maximum $2\%$ difference was found between the power at different polarisations after the lens. Line scanning was performed by a piezo stage moving at 20 \textmu m/s, with the centre of the laser spot visually aligned with the centre of the nanoisland row (Figure \ref{fig1} b)). For the statistics gathered in Figure \ref{fig3} c-d), nanoislands that were induced into a single-vortex state, or unidentifiable spin textures, were not counted towards double-vortex fidelities. All nanoislands were subsequently re-saturated with a permanent magnet and measured via MFM to verify that no clear damage to the nanoislands had occurred from the laser illumination process. Nanoislands displaying signs of moderate damage were excluded from the total statistics. Laser polarisations of ±36$^{\circ}$ and ±54$^{\circ}$ were only performed on Array 1 in limited number and are not shown in the figure (full data is available in the Supporting Information Tables S1 and S2). Laser-writing for these statistics was performed in both the +X and -X directions. Arrays were initially saturated in both the positive and negative X directions, with resulting statistics combined regardless of scan direction or initial saturation direction. Each illuminated row is recorded as a single data point for each $\pm$45$^{\circ}$ nanoisland orientation. Average fidelities and standard deviations are reported.
\subsection*{Tip-writing}
Magnetic tip-writing was performed on a Veeco Dimension 3100 Magnetic Force Microscope in tapping mode, using a commercially-available high-moment tip (Bruker MESP-HM), at a low scanning speed ($\approx$ 1 to 2 \textmu\space m/s), in an orientation shown by Figure \ref{fig1} b). The exact location of the tip with respect to the nanoislands is manually controlled to $\approx\pm$ 500 nm, more precise tip location can be discerned from resulting vortex locations. Measurements were analysed with Gwyddion \cite{gwyddion}.
\subsection*{MFM Imaging}
Samples were magnetically imaged on the same system, using a low (Bruker MESP-LM-V2) or normal moment (Bruker MESP-V2) tip. Magnetic fields were applied to the sample via an electromagnet with saturation performed with a strong permanent magnet. Coercive and saturation field values were determined using a commercial Durham Magneto Optics nanoMOKE2 system to measure the longitudinal Magneto-Optical Kerr Effect (`MOKE') signal. 
\subsection*{Simulations and plotting}
Absorption simulations were performed with the Lumerical Inc. FDTD solver \cite{lumerical}. Micromagnetic simulations were performed with MuMax3 \cite{mumaxa,mumaxb}. Saturation magnetization was set to $\mathrm{M_{sat}}$ = 750 $\times 10^{3}$ A/m, exchange stiffness $\mathrm{A_{ex}}$ = 13 $\times 10^{-12}$, and magnetization with a stopping condition of 1 $\times 10^{-6}$. Hysteresis simulations were performed statically, with energy minimisation at each field step, with field applied at an angle of $\pm$1$^{\circ}$ to the 45$^{\circ}$ direction (to break symmetry) and then calculated switching fields averaged. The applied magnetic field was incremented by +0.1 mT between each energy minimization step until reversal was observed. Single-vortex switching fields were averaged across both angular nanoisland orientations and both chiralities. The averaging was performed on multiple initial random seeds. The magnetic simulation used to calculate the local stray dipolar field reduction was implemented with $\mathrm{M_{sat}}$ = 800 $\times 10^{3}$ A/m, matching the geometry of the L.A. Array, and with no stray dipolar field from the vortex texture. Hysteresis simulations were discretised into cells of 2 nm $\times$ 2 nm $\times$ 5 nm, with other simulations discretised into 2 $\times$ nm 2 $\times$ nm 20 $\times$ nm. Dimensions were chosen to match the measured array geometries as closely as possible while maintaining an even number of cells, and with highest prime factors of 7 or less where possible, for optimal performance \cite{mumaxa}. Figure \ref{fig5} e,h) was created with MATLAB\cite{matlab}. All other analysis and plotting was performed with python\cite{python}, numpy\cite{numpy} and matplotlib\cite{matplotlib}.

\subsection*{Acknowledgements}
H.H. was supported by the EPSRC DTP award EP/T51780X/1 and EPSRC Doctoral Prize Fellowship EP/W524323/1. W.B, R.O and D.B. were supported by the Imperial President's Excellence Fund for Frontier Research J.C.G. was supported by a Royal Academy of Engineering Research Fellowship and EPSRC grant EP/Y003276/1. T.D. is supported by International Research Fellow of Japan Society for the Promotion of Science (Postdoctoral Fellowships for Research in Japan) JSPS KAKENHI Grant No. 21F20790. J.C.G. and W.B. were supported by EPSRC grant EP/X015661/1. K.D.S. was supported by The Eric and Wendy Schmidt Fellowship Program and the Engineering and Physical Sciences Research Council (Grant No. EP/W524335/1). A.V. was supported by EPSRC IAA funding.

\subsection*{Author contributions}
H.H. and J.C.G. conceived the work.\\
H.H. drafted the manuscript, with contributions from all authors in editing and revision stages.\\
J.C.G. fabricated the ASVI.\\
H.H. performed the laser-writing of horizontally saturated arrays and applied magnetic field protocols.\\
T.Z., H.H., T.F., and D.B. performed the preliminary laser-writing tests of vertically saturated and demagnetised arrays.\\
J.C.G. and H.H. performed the tip-writing.\\
H.H., J.C.G., T.Z., T.F., and D.B. performed MFM measurements.\\
A.V. performed the simulation for the calculation and analysis of the stray dipolar field in Figure \ref{fig5} i).\\
T.D. created the programs for the calculation and analysis of the spatial demagnetising field landscape in Figure \ref{fig5} c-g) and Supporting Information Figures S17-18.\\
H.H. performed all other simulations, with contributions from K.D.S. and T.D.\\
T.Z., J.C.G., A.V., and H.H. performed MOKE measurements of coercive field.\\
X.X. performed absorption simulations and measurements of laser spot diameter.\\
A.V. and T.Z. performed SEM measurements and analysis of nanoisland dimensions. The plots in Supplementary Figures S1-4 were created by A.V.\\
W.B. and R.O. kept the research on track.\\
Simulations were performed on the Imperial College Research Computing Service \cite{hpc}.\\
The authors would like to thank David Mack for excellent laboratory management.\\
The authors would like to thank the groups of R.O. and W.B. for their helpful discussions.

\newpage
\renewcommand{\thefigure}{S\arabic{figure}}
\section*{Supporting Information for ``Magnetic vortex writing and local reversal seeding in artificial spin-vortex ice via all-optical and surface-probe control''}

Supplementary material for the paper ``Magnetic vortex writing and local reversal seeding in artificial spin-vortex ice via all-optical and surface-probe control", showing double-vortices induced without narrow bars, MuMax3\cite{mumaxa,mumaxb} micromagnetic simulations of single wide bars reversal pathways, modification of the local field landscape by neighbouring vortices, simulated Lumerical FDTD\cite{lumerical} absorption data, and details about applied global magnetic fields. \\

\subsection*{Array dimensions}

\begin{figure*}[h]
    \includegraphics[width=\textwidth]{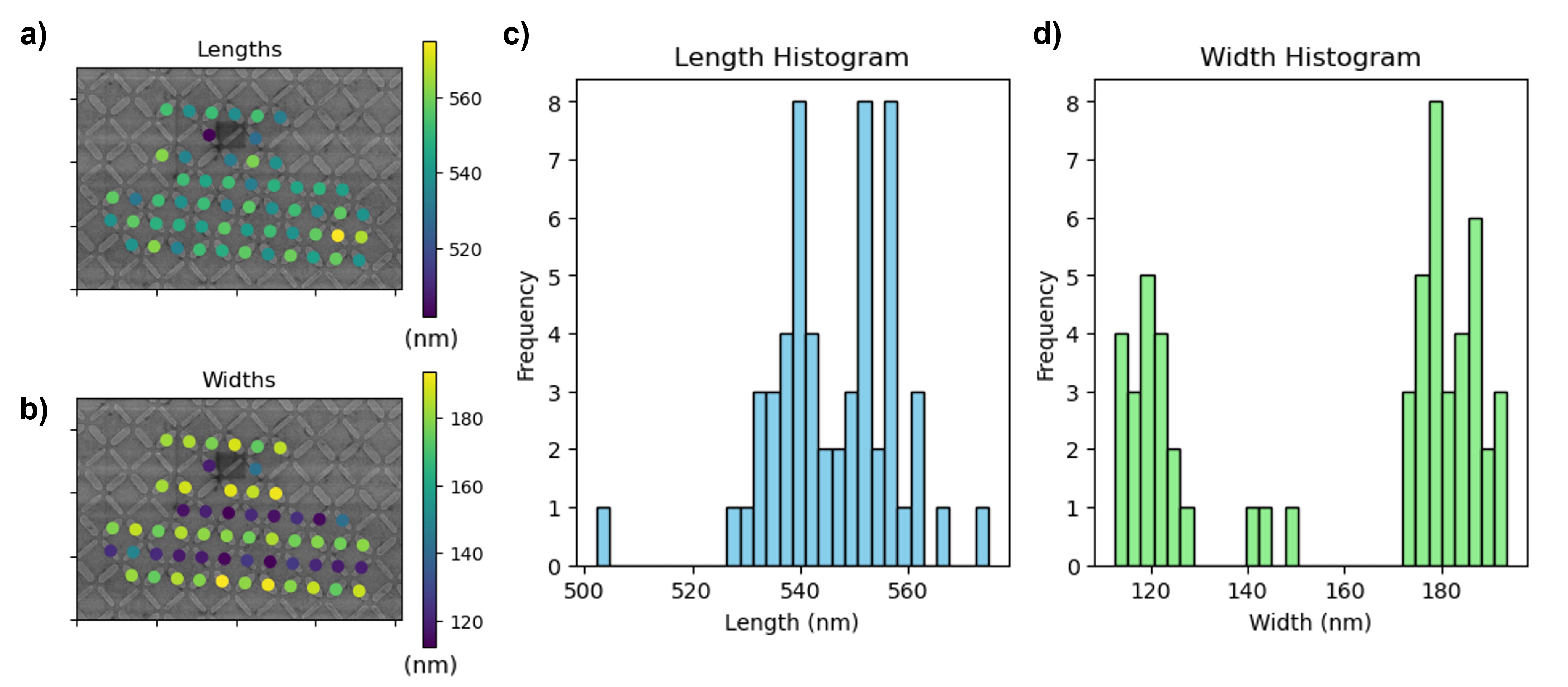}
\caption{\label{figS1} \textbf{a-b)} Reference images showing measured lengths and widths of sample bars in Array 1. \textbf{c-d)} Corresponding histograms of bar lengths and widths.}
\end{figure*}
\begin{figure*}
    \includegraphics[width=\textwidth]{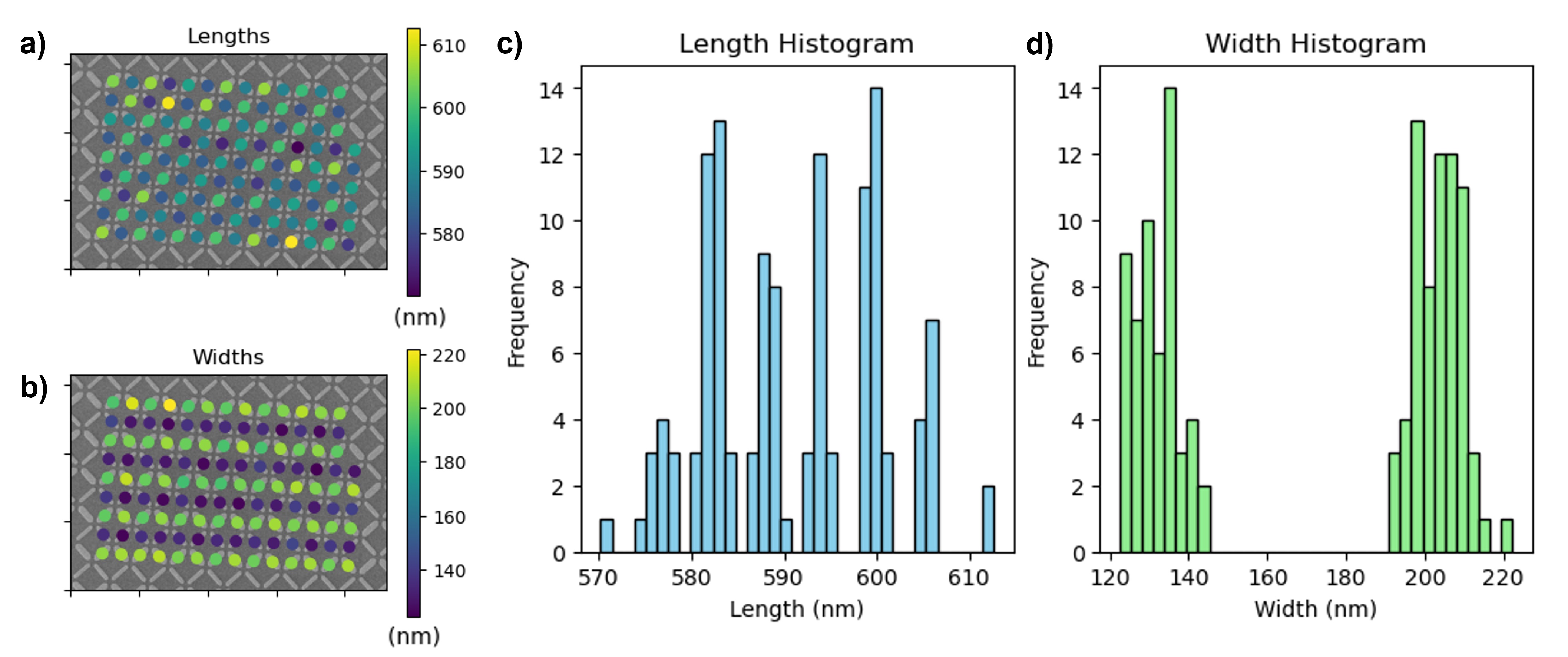}
\caption{\label{figS2} \textbf{a-b)} Reference images showing measured lengths and widths of sample bars in Array 2. \textbf{c-d)} Corresponding histograms of bar lengths and widths.}
\end{figure*}
\begin{figure*}
    \includegraphics[width=\textwidth]{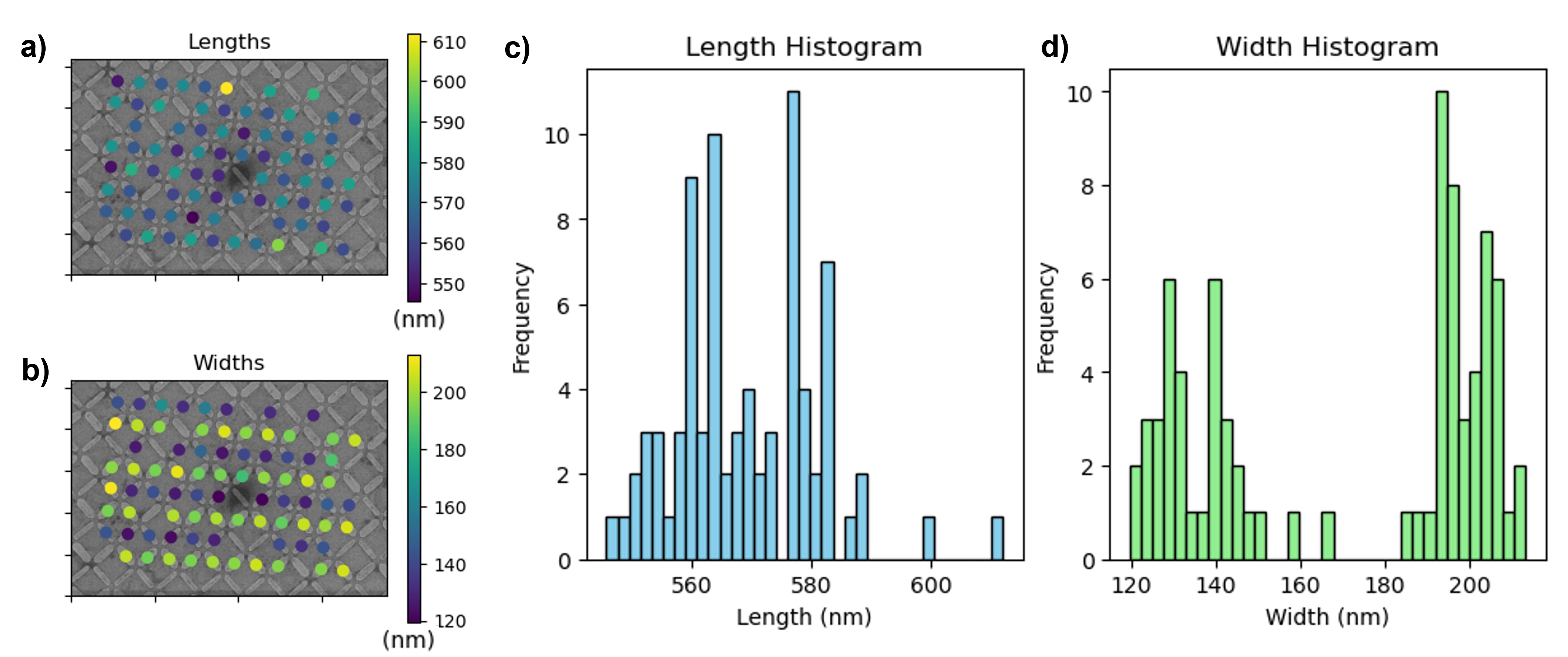}
\caption{\label{figS3} \textbf{a-b)} Reference images showing measured lengths and widths of sample bars in Array 3. \textbf{c-d)} Corresponding histograms of bar lengths and widths.}
\end{figure*}
\begin{figure*}
    \includegraphics[width=\textwidth]{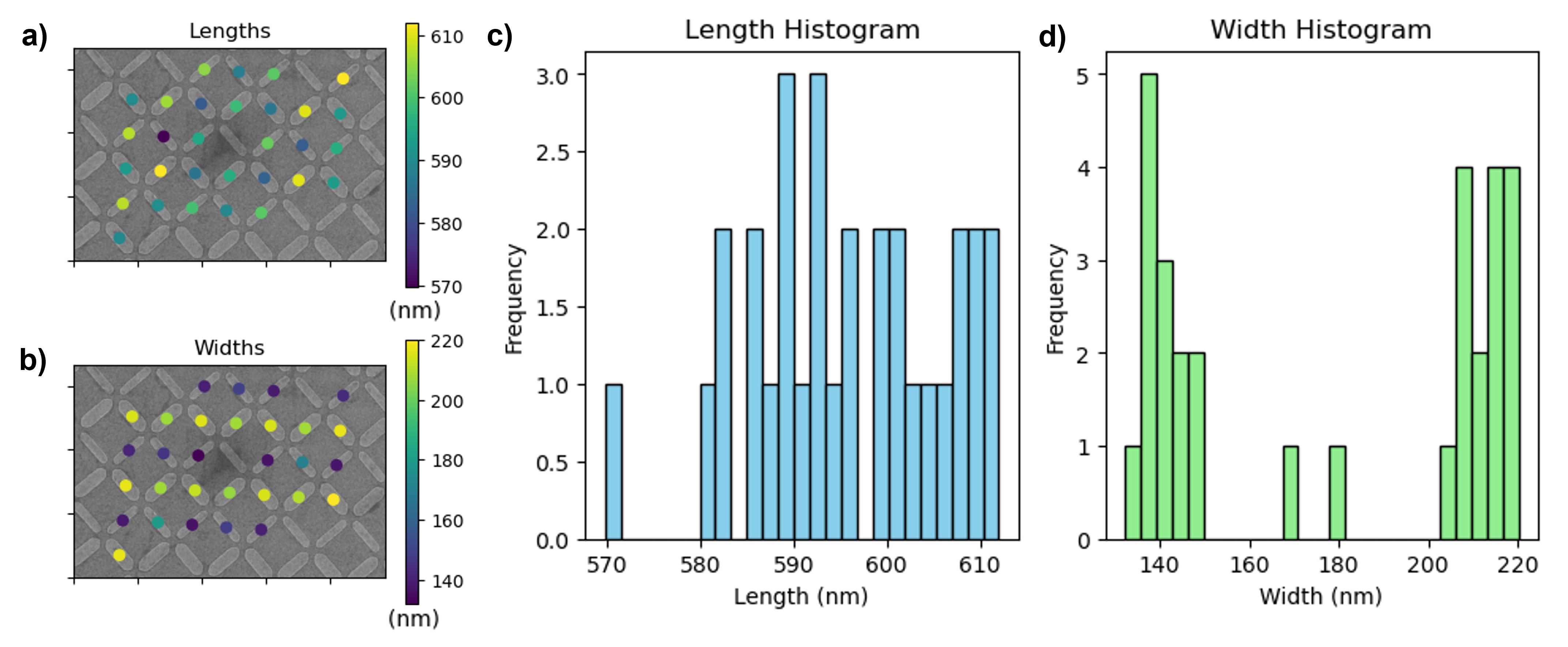}
\caption{\label{figS4} \textbf{a-b)} Reference scanning electron microscopy images showing measured lengths and widths of sample bars in Array 4. \textbf{c-d)} Corresponding histograms of bar lengths and widths.}
\end{figure*}

Figures \ref{figS1}-\ref{figS4} show the sampled bars used to determine the lengths and widths of wide and narrow bars within each 26$\times$26 array, alongside the corresponding histograms. Images were taken via scanning electron microscopy.

\subsection*{Surface probe vortex tip-writing and chirality control}
\begin{figure*}
    \includegraphics[width=\textwidth]{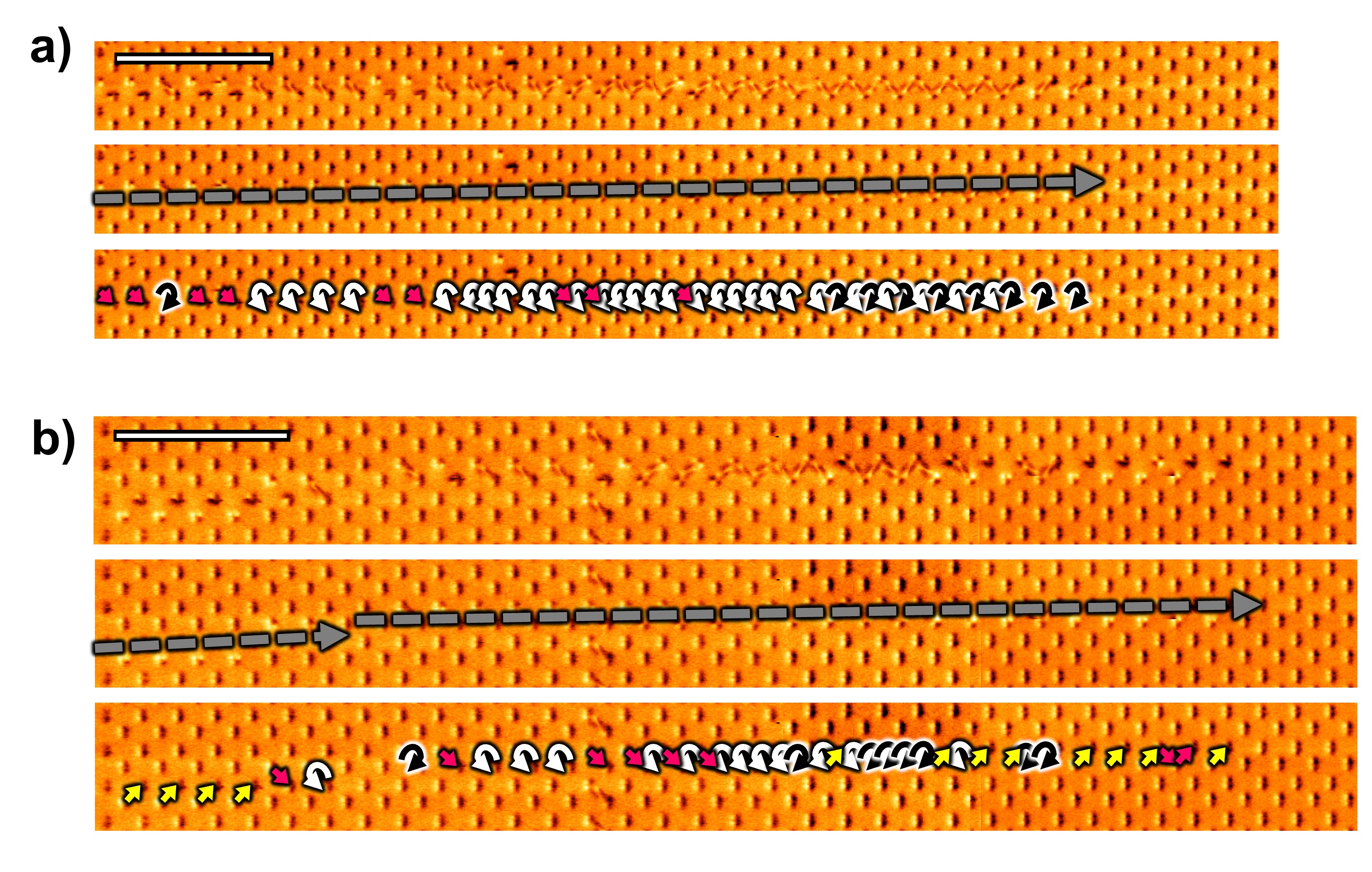}
\caption{\label{figS5} \textbf{a)} Triplicated image of tip-written line of single-vortices and macrospin-to-macrospin flips, with the first image showing an MFM collage of the row, the second image showing the approximate deduced path of the high-moment tip with the grey dotted arrow, and the third showing annotated vortex chirality and reversed macrospin orientations. The scale bar shows 5 \textmu m. \textbf{b)} Triplicated image of another tip-written row. The approximate deduced initial irregular motion of the high-moment tip can be seen from the change in angle between the grey dotted arrows. Unlabelled vortices and macrospin reversals were induced by a different tip-write. The images are a collage of different MFM scans.}
\end{figure*}

\begin{figure*}
    \includegraphics[width=\textwidth]{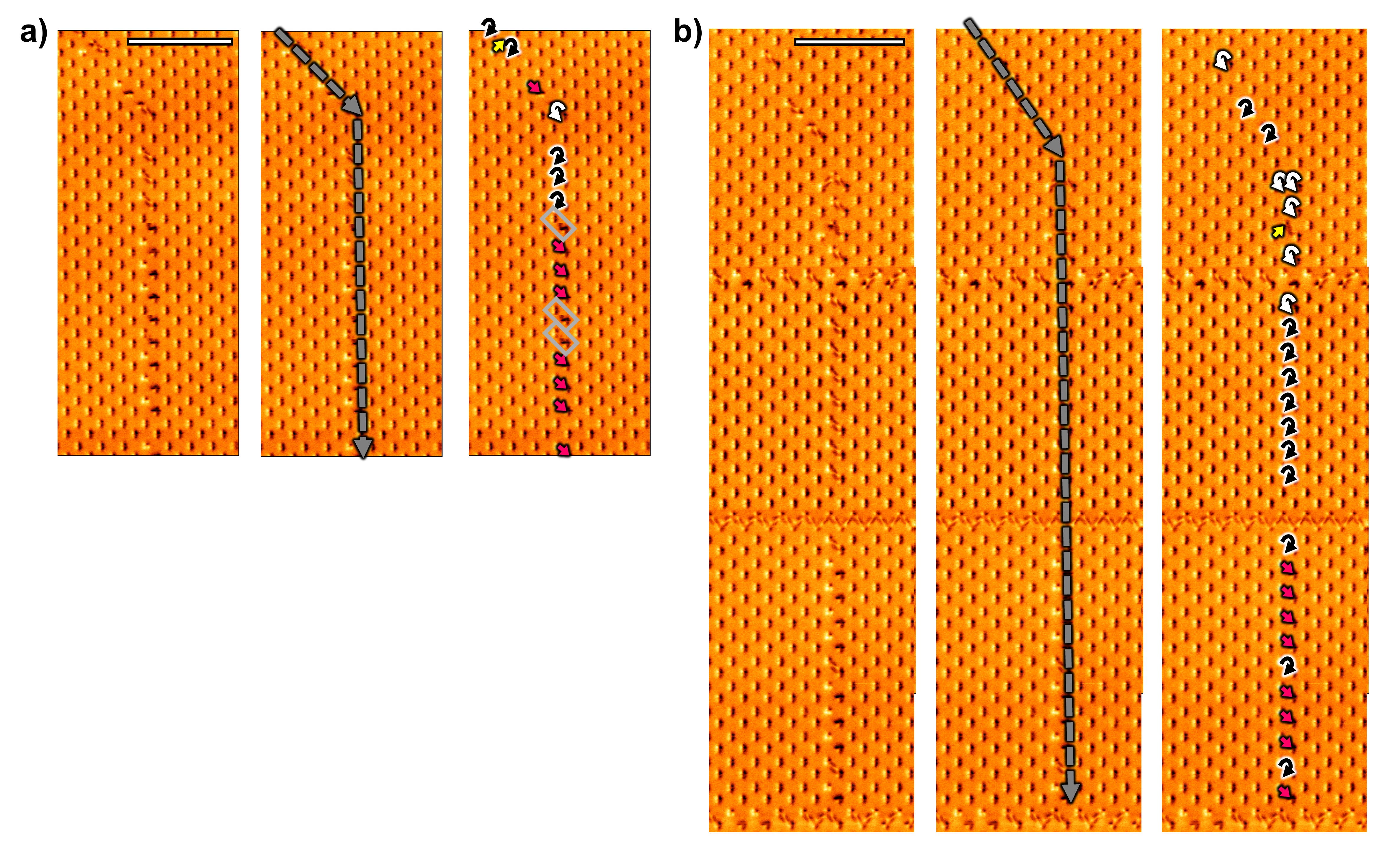}
\caption{\label{figS6} \textbf{a)} Triplicated image of vertical tip-written line of single-vortices and macrospin-to-macrospin flips, with the first image showing an MFM collage of the row, the second image showing the approximate deduced path of the high-moment tip with the grey dotted arrow, and the third showing annotated vortex chirality and reversed macrospin orientations. The deduced initial irregular motion of the high-moment tip can be seen from the change in angle between the grey dotted arrows. The scale bar shows 5 \textmu m. \textbf{b)} Triplicated image of another tip-written row. Unlabelled vortices and macrospin reversals were induced by different tip-writes. A narrow nanomagnet has been induced into a single-vortex state, and another narrow bar has reversed macrospin direction. The images are a collage of different MFM scans.}
\end{figure*}

Figure \ref{figS5} shows triplicated images of two horizontal tip writes, with macrospin-to-macrospin (`M2M') reversals and induced single-vortices labelled. Black circular arrows annotate clockwise vortices, while white arrows annotate anti-clockwise vortices. M2M reversals are labelled as -45$^{\circ}$ bars with south-east macrospins in pink, +45$^{\circ}$ bars with north-east in yellow. Unlabelled vortices were induced by different tip writes.

The approximate path of the tips were deduced from the locations of induced vortices and macrospin reversals, and where multiple rows were crossed in a single write. These examples suggest that M2M bar orientations and induced vortex chirality are influenced by the location of the tip with respect to the nanomagnet.

Similarly, Figure \ref{figS6} shows duplicated images of two vertical tip-writes, with the tip moving from up to down, performed at the same relative angle to the vertical as those in Figure \ref{figS5} were to the horizontal. Unlabelled vortices were induced by a different, horizontal tip write. Vertical tip-writing induces vortices in alternating rows, due to the alternating pattern of wide and narrow nanobars. The distribution of vortices at the top of the images suggests that the tip may have jumped slightly at the start of the writes.

\subsection*{Optically written double-vortices and Experimentally customised switching fields}

\begin{table}
    \begin{tabular}{|p{0.04\textwidth}|p{0.068\textwidth}|p{0.068\textwidth}|p{0.068\textwidth}|p{0.068\textwidth}|p{0.068\textwidth}|p{0.065\textwidth}|p{0.065\textwidth}|p{0.068\textwidth}|p{0.068\textwidth}|p{0.068\textwidth}|}\hline
         \textbf{Array 1} & \textbf{+45 D.V.} & \textbf{-45 D.V.} & \textbf{+45 S.V.} & \textbf{-45 S.V.} & \textbf{+45 O.T.} & \textbf{-45 O.T.} & \textbf{+45 Dam} & \textbf{-45 Dam} & \textbf{+45 T.U.} & \textbf{-45 T.U.}\\ \hline
         \textbf{0$^{\circ}$} & 0, 0, 0, 0 &  0, 0, 0, 0 & 0, 0, 0, 0 &  0, 0, 0, 0 & 0, 0, 0, 0 & 0, 0, 0, 0 & 0, 0, 0, 0 &  0, 0, 0, 0 & 13, 13, 13, 13 & 11, 11, 12, 12\\ \hline
         \textbf{30$^{\circ}$}	& 0, 0, 0, 0, 0, 0 & 2, 4, 3, 7, 8, 10 & 0, 0, 0, 0, 0, 0 & 0, 0, 0, 0, 1, 0 & 0, 1, 0, 0, 0, 0 & 0, 0, 0, 0, 1, 0 & 0, 0, 0, 0, 0, 0 & 0, 0, 0, 0, 0, 0 & 13, 13, 13, 13, 13, 13 & 13, 12, 13, 13, 12, 13\\ \hline
         \textbf{36$^{\circ}$} & 0, 0, 0 & 5, 7, 6 & 0, 0, 0 & 0, 0, 0 & 0, 1, 0 & 0, 0, 0 & 0, 0, 0 & 0, 0, 0 & 13, 13, 13 & 13, 11, 13 \\ \hline
         \textbf{45$^{\circ}$} & 0, 0, 4,  0, 0, 0, 4, 1, 0, 1, 3, 1 & 3, 3, 10, 5, 11, 8, 12, 12, 5, 11, 9, 8 & 0, 0, 0, 0, 0, 0, 0, 0, 0, 0, 0, 0 & 0, 0, 0, 0, 0, 0, 0, 0, 0, 0, 0, 0 & 0, 0, 0, 0, 0, 0, 0, 2, 0, 0, 0, 0 & 0, 0, 0, 0, 0, 0, 0, 0, 1, 0, 0, 0 & 0, 0, 0, 0, 0, 0, 0, 0, 0, 0, 0, 0 & 0, 0, 0, 0, 0, 0, 0, 0, 0, 0, 0, 0 & 13, 13, 13, 13, 13, 13, 13, 13, 13, 13, 13, 13 & 13, 13, 13, 13, 13, 13, 12, 13, 13, 13, 13, 13\\ \hline
         \textbf{54$^{\circ}$} & 4, 4, 0 & 5, 10, 5 & 0, 0, 0 & 0, 0, 0 & 0, 0, 0 & 0, 0, 0 & 0, 0, 0 & 0, 0, 0  & 13, 13, 13 & 12, 13, 13 \\ \hline
         \textbf{60$^{\circ}$} & 0, 2, 4, 0, 1, 1 & 0, 1, 4, 6, 6, 7 & 0, 0, 0, 0, 0, 0 & 0, 0, 0, 0, 0, 0 & 0, 0, 0, 0, 1, 0 & 0, 0, 0, 0, 1, 0 & 0, 0, 0, 0, 0, 0 & 0, 0, 0, 0, 0, 0 & 13, 13, 13, 13, 13, 13 & 13, 13, 13, 13, 13, 13\\ \hline
         \textbf{90$^{\circ}$} & 0, 5, 2, 0 & 0, 0, 1, 0 & 0, 0, 0, 0 & 0, 0, 0, 0 & 0, 0, 0, 0 & 0, 0, 0, 0 & 0, 0, 0, 0 & 2, 0, 0, 0 & 13, 13, 13, 13 & 11, 12, 12, 11\\ \hline
         \textbf{-60$^{\circ}$} & 12, 12, 8, 5, 5, 0 & 0, 0, 0, 0, 0, 0 & 1, 0, 0, 0, 0, 0 & 0, 0, 0, 0, 0, 0 & 0, 1, 0, 0, 0, 0 & 0, 0, 0, 0, 0, 0 & 0, 0, 0, 0, 0, 0 & 0, 1, 0, 0, 0, 0 & 13, 13, 13, 13, 13, 13 & 13, 12, 13, 13, 12, 13\\ \hline
         \textbf{-54$^{\circ}$} & 3, 10, 5 & 0, 1, 1 & 1, 0, 0 & 0, 0, 0 & 0, 0, 0 & 0, 0, 0 & 0, 0, 0 & 0, 0, 0 & 13, 13, 13 & 13, 13, 13\\ \hline
         \textbf{-45$^{\circ}$} & 13, 10, 13, 2, 0, 10, 11, 9, 0, 13, 10, 12 & 0, 0, 0, 0, 0, 1, 0, 0, 0, 0, 0, 0 & 0, 2, 0, 0, 0, 0, 0, 0, 1, 0, 0, 0 & 0, 0, 0, 0, 0, 0, 0, 0, 0, 1, 0, 0 & 0, 0, 0, 0, 0, 0, 0, 0, 0, 0, 0, 0 & 0, 0, 0, 0, 0, 0, 0, 0, 0, 0, 0, 0 & 0, 0, 0, 0, 0, 0, 0, 0, 0, 0, 0, 0 & 0, 0, 0, 0, 0, 0, 0, 0, 0, 0, 0, 0 & 13, 13, 13, 13, 13, 13, 13, 13, 13, 13, 13, 13 & 13, 13, 13, 13, 13, 13, 13, 11, 13, 13, 13, 12\\ \hline
         \textbf{-36$^{\circ}$} & 0, 0, 1 & 0, 1, 0 & 0, 0, 0 & 0, 0, 0 & 0, 0, 0 & 0, 0, 0 & 0, 0, 0 & 0, 0, 0 & 13, 13, 13 & 13, 13, 12\\ \hline
         \textbf{-30$^{\circ}$} & 3, 12, 3, 0, 3, 4 & 0, 0, 0, 0, 1, 0 & 1, 0, 0, 0, 0, 0 & 0, 0, 0, 0, 0, 0 & 0, 0, 0, 0, 0, 0 & 0, 0, 0, 0, 0, 0 & 0, 0, 0, 0, 0, 0 & 0, 0, 0, 0, 0, 0 & 13, 13, 13, 13, 13, 13 & 13, 13, 13, 13, 13, 13\\ \hline
    \end{tabular}
    \caption{Table showing number of wide bars in each row by bar orientation ($\pm$45$^{\circ}$) in Array 1 after laser illumination by state (D.V. = double-vortex, S.V. = single-vortex, O.T. = Other domain texture/unidentified texture, Dam = Newly Damaged) with the final columns showing the total bars in each row that were undamaged (T.U. = Total Undamaged). Table rows are separated by laser polarisation, while commas separate different array row illuminations.}
    \label{table1}
\end{table}

\begin{table}
    \begin{tabular}{|p{0.04\textwidth}|p{0.068\textwidth}|p{0.068\textwidth}|p{0.068\textwidth}|p{0.068\textwidth}|p{0.068\textwidth}|p{0.065\textwidth}|p{0.065\textwidth}|p{0.068\textwidth}|p{0.068\textwidth}|p{0.068\textwidth}|}\hline
         \textbf{Array 3} & \textbf{+45 D.V.} & \textbf{-45 D.V.} & \textbf{+45 S.V.} & \textbf{-45 S.V.} & \textbf{+45 O.T.} & \textbf{-45 O.T.} & \textbf{+45 Dam} & \textbf{-45 Dam} & \textbf{+45 T.U.} & \textbf{-45 T.U.}\\ \hline
         \textbf{0$^{\circ}$} & 0, 0, 0, 0 & 2, 0, 1, 1 & 0, 0, 0, 0 & 0, 0, 0, 0 & 0, 0, 0, 0 & 0, 0, 0, 0 & 0, 0, 0, 0 & 0, 0, 0, 0 &13, 13, 13, 13 & 12, 12, 13, 13\\ \hline
         \textbf{30$^{\circ}$}	& 0, 0, 0, 0, 0, 0 & 9, 8, 9, 0, 0, 0 & 0, 0, 0, 0, 0, 0 & 0, 0, 0, 0, 0, 0 & 0, 0, 0, 0, 0, 0 & 0, 0, 0, 0, 0, 0 & 0, 0, 0, 0, 0, 0 & 0, 0, 0, 0, 0, 0 & 13, 13, 13, 13, 13, 13 & 13, 11, 13, 13, 11, 13\\ \hline
         \textbf{45$^{\circ}$} & 0, 0, 3, 0, 0, 0, 0, 0, 0, 1, 5, 2 & 6, 11, 12, 0, 1, 0, 4, 6, 5, 13, 12, 12 & 0, 0, 0, 0, 0, 0, 0, 0, 0, 1, 0, 0 & 0, 0, 0, 0, 0, 0, 0, 0, 0, 0, 0, 0 & 0, 0, 0, 0, 0, 0, 0, 0, 0, 0, 0, 0 & 1, 0, 1, 0, 0, 0, 0, 0, 0, 0, 0, 0 & 0, 0, 0, 0, 0, 0, 0, 0, 0, 0, 1, 0 & 1, 0, 0, 1, 0, 1, 1, 0, 0, 0, 0, 0 & 13, 13, 13, 13, 13, 13, 13, 13, 13, 13, 12, 13 & 12, 13, 13, 11, 13, 12, 10, 13, 11, 13, 12, 13\\ \hline
         \textbf{60$^{\circ}$} & 1, 4, 1, 1, 0, 1 & 11, 12, 9, 2, 1, 2 & 0, 0, 0, 0, 0, 0 & 0, 0, 0, 0, 0, 0 & 0, 0, 0, 0, 0, 0 & 0, 1, 1, 0, 0, 0 & 0, 0, 0, 0, 0, 0 & 0, 0, 0, 0, 0, 0 & 13, 13, 13, 13, 13, 13 & 13, 13, 13, 13, 13, 13\\ \hline
         \textbf{90$^{\circ}$} & 1, 12, 1, 1 & 0, 7, 2, 0 & 0, 0, 0, 0 & 0, 0, 0, 0 & 0, 0, 0, 0 & 0, 0, 0, 1 & 0, 0, 0, 0 & 1, 0, 0, 1 & 13, 13, 13, 13 & 12, 13, 12, 12\\ \hline
         \textbf{-60$^{\circ}$} & 13, 6, 2, 1, 0, 0 & 0, 0, 0, 0, 0, 0 & 0, 0, 0, 0, 0, 0 & 0, 0, 0, 0, 0, 0 & 0, 0, 0, 0, 0, 0 & 0, 0, 0, 0, 0, 0 & 0, 0, 0, 0, 0, 0 & 0, 1, 0, 0, 0, 0 & 13, 13, 13, 13, 13, 13 & 13, 11, 13, 13, 11, 13\\ \hline
         \textbf{-45$^{\circ}$} & 13, 13, 12, 0, 0, 0, 1, 3, 7, 8, 9, 8 & 0, 0, 0, 0, 0, 0, 0, 0, 0, 0, 0, 0 & 0, 0, 0, 0, 0, 0, 0, 0, 0, 0, 0, 0 & 0, 0, 0, 0, 0, 0, 0, 0, 0, 1, 0, 0 & 0, 0, 1, 0, 0, 0, 0, 0, 1, 0, 0, 0 & 0, 0, 0, 0, 0, 0, 0, 0, 0, 0, 0, 1 & 0, 0, 0, 0, 0, 0, 0, 0, 0, 0, 0, 0 & 0, 0, 0, 0, 0, 0, 0, 0, 0, 0, 0, 0 & 13, 13, 13, 13, 13, 13, 13, 13, 13, 13, 13, 13 & 13, 13, 13, 12, 13, 13, 13, 12, 13, 13, 13, 12\\ \hline
         \textbf{-30$^{\circ}$} & 3, 2, 4, 0, 3, 1 & 0, 0, 0, 0, 0, 0 & 0, 0, 0, 0, 0, 0 & 0, 0, 0, 0, 0, 0 & 0, 0, 0, 0, 0, 0 & 0, 0, 0, 0, 0, 0 & 0, 0, 0, 0, 0, 0 & 0, 0, 0, 0, 0, 0 & 13, 13, 13, 13, 13, 13 & 13, 13, 13, 13, 13, 13\\ \hline    
    \end{tabular}
    \caption{Table showing number of wide bars in each row by bar orientation ($\pm$45$^{\circ}$) in Array 3 after laser illumination by state (D.V. = double-vortex, S.V. = single-vortex, O.T. = Other domain texture/unidentified texture, Dam = Newly Damaged) with the final columns showing the total bars in each row that were undamaged (T.U. = Total Undamaged). Table rows are separated by laser polarisation, while commas separate different array row illuminations.}
    \label{table2}
\end{table}

Tables \ref{table1} and \ref{table2} show the raw data used to calculate the optical double-vortex writing fidelities of Arrays 1 and 3 respectively. Columns show the number of double-vortices (`D.V.'), single-vortices (`S.V.'), other unidentifiable multi-domain textures (`O.T.'), newly damaged bars (`Dam'), and the total number of undamaged bars (`T.U.') in each illuminated array row, with all categories split into bar orientations of +45$^{\circ}$ and -45$^{\circ}$. The percentages shown in Figure 2 of the main text were calculated by dividing the number of optically induced double-vortices by the the number of undamaged total bars, for each $\pm45^{\circ}$ orientation. Table rows show the statistics for different laser polarisations. Damage was assessed from magnetic re-saturation of the array, with bars categorised as `damaged' if there were signs of strong residual multi-domain textures remaining.
   
\begin{figure*}[t!]
\centering
\includegraphics[width=1.0\textwidth]{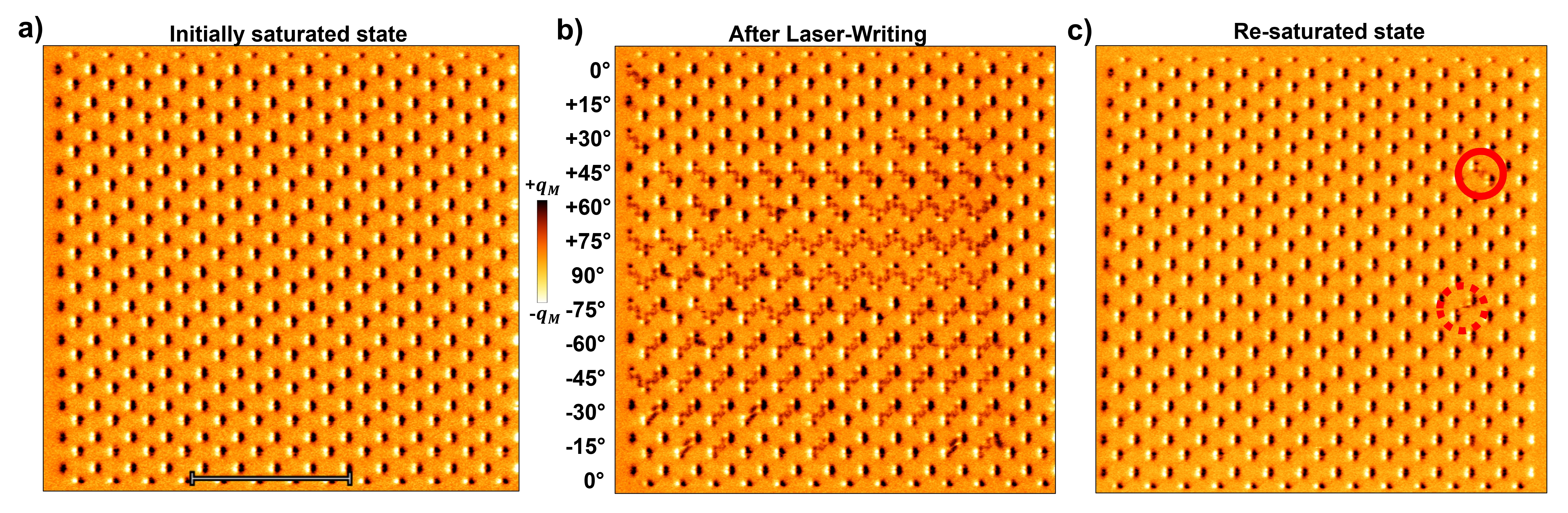}
\caption{\label{figS7} \textbf{a)} MFM image showing the initially saturated state of Array 4. Scale bar = 5 \textmu m. \textbf{b)} MFM image showing the same array after laser-writing of all wide bar rows (P = 5.0 mW). The laser polarisation is labelled on the left hand side of each row. Each laser line illumination ended before reaching the end of the row. \textbf{c)} MFM image showing the subsequent re-saturated state of the array. Damaged bars are circled in red (solid circle clear remnant multi-domain texture, dashed circle partial damage).}
\end{figure*}

Figure \ref{figS7} shows the initially saturated (a)), laser-written (b)), and re-saturated state (c)) of Array 4, as seen in Figure 3 c-d) of the main text. Damaged bars showing remnant multi-domain textures even after re-saturation are circled in red.

\begin{figure*}
    \includegraphics[width=\textwidth]{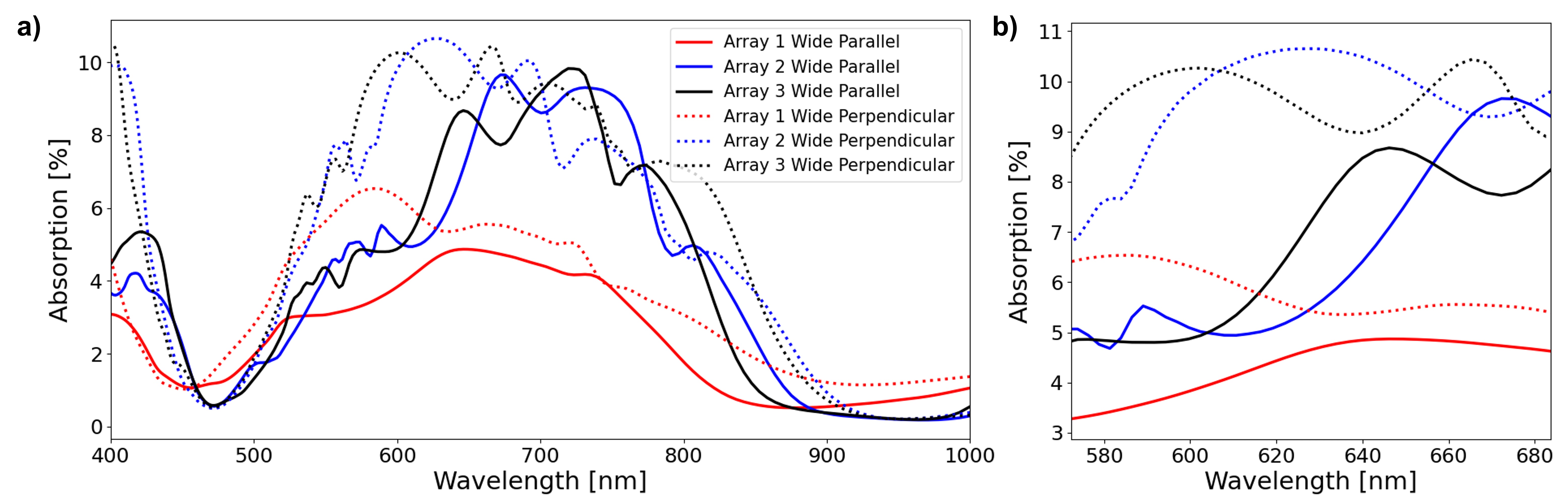}
\caption{\label{figS8} \textbf{a)} Simulated absorption within the wide nanoislands as a function of wavelength for Arrays 1, 2, and 3. `Perpendicular' and `Parallel' denote the orientation of the laser polarisation with respect to the long axis of the nanoisland bar. \textbf{b)} Zoomed-in section of a), highlighting the region around the laser wavelength, 633 nm.}
\end{figure*}

Figure \ref{figS8} a) and b) show the simulated absorption within the wide nanoislands as a function of wavelength for Arrays 1, 2, and 3, calculated using the Lumerical FDTD Solver\cite{lumerical}. Absorption was calculated for both Perpendicular and Parallel orientations of the linear laser polarisation with respect to the bar orientation. Due to the nature of initialising the unit cell within the simulation, the lengths of wide and narrow nanoislands in Array 2 were averaged.

\begin{figure*}
    \includegraphics[width=\textwidth]{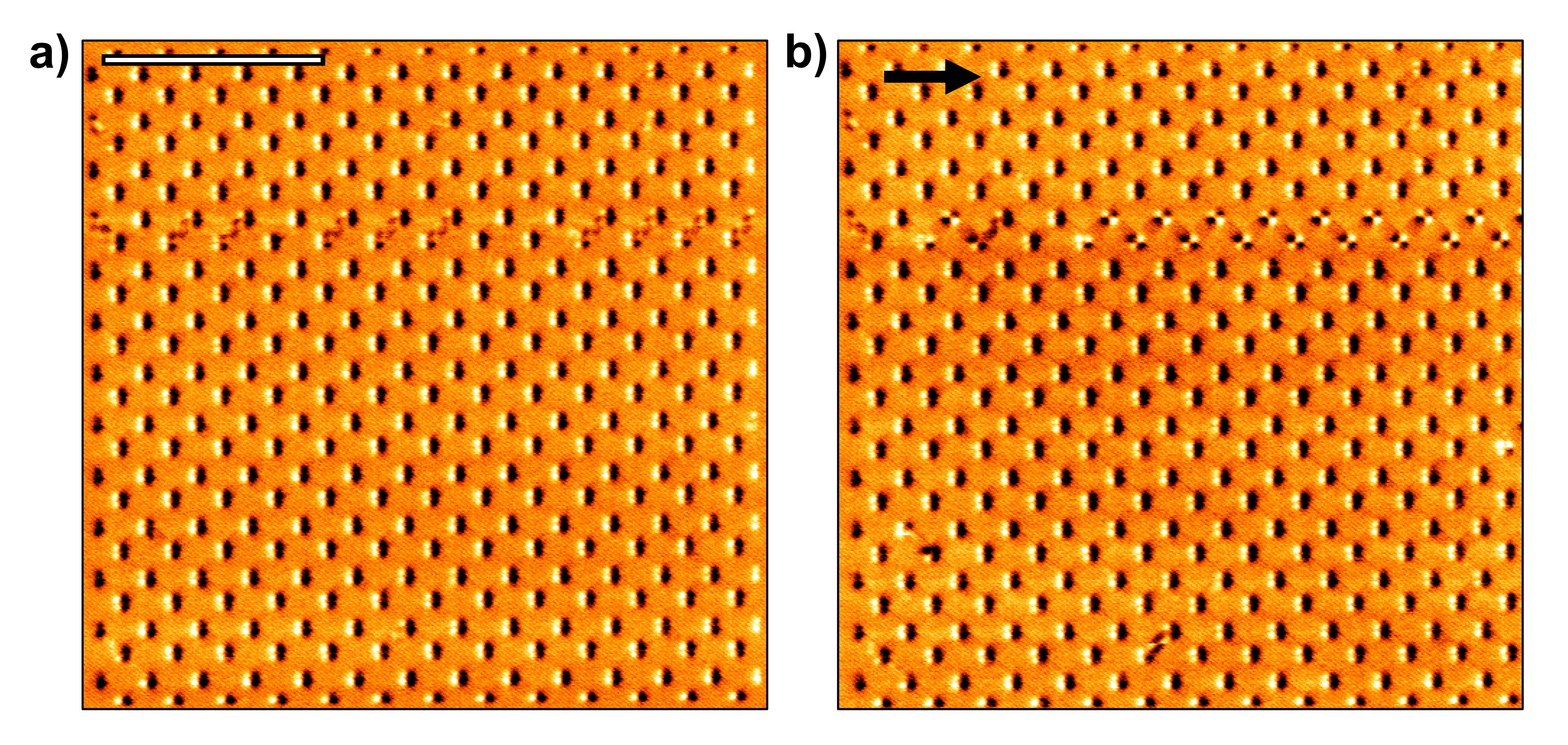}
\caption{\label{figS9} \textbf{a)} MFM image of initially -X saturated Array 1 after laser-writing has induced several double-vortices in a wide bar row. The scale bar is 5 \textmu m long. \textbf{b)} Image of the subsequent microstate of the array after application of a magnetic field of magnitude 91\% H\textsubscript{c-start} in the +X direction, as indicated by the arrow.}
\end{figure*}

\begin{figure*}
    \includegraphics[width=\textwidth]{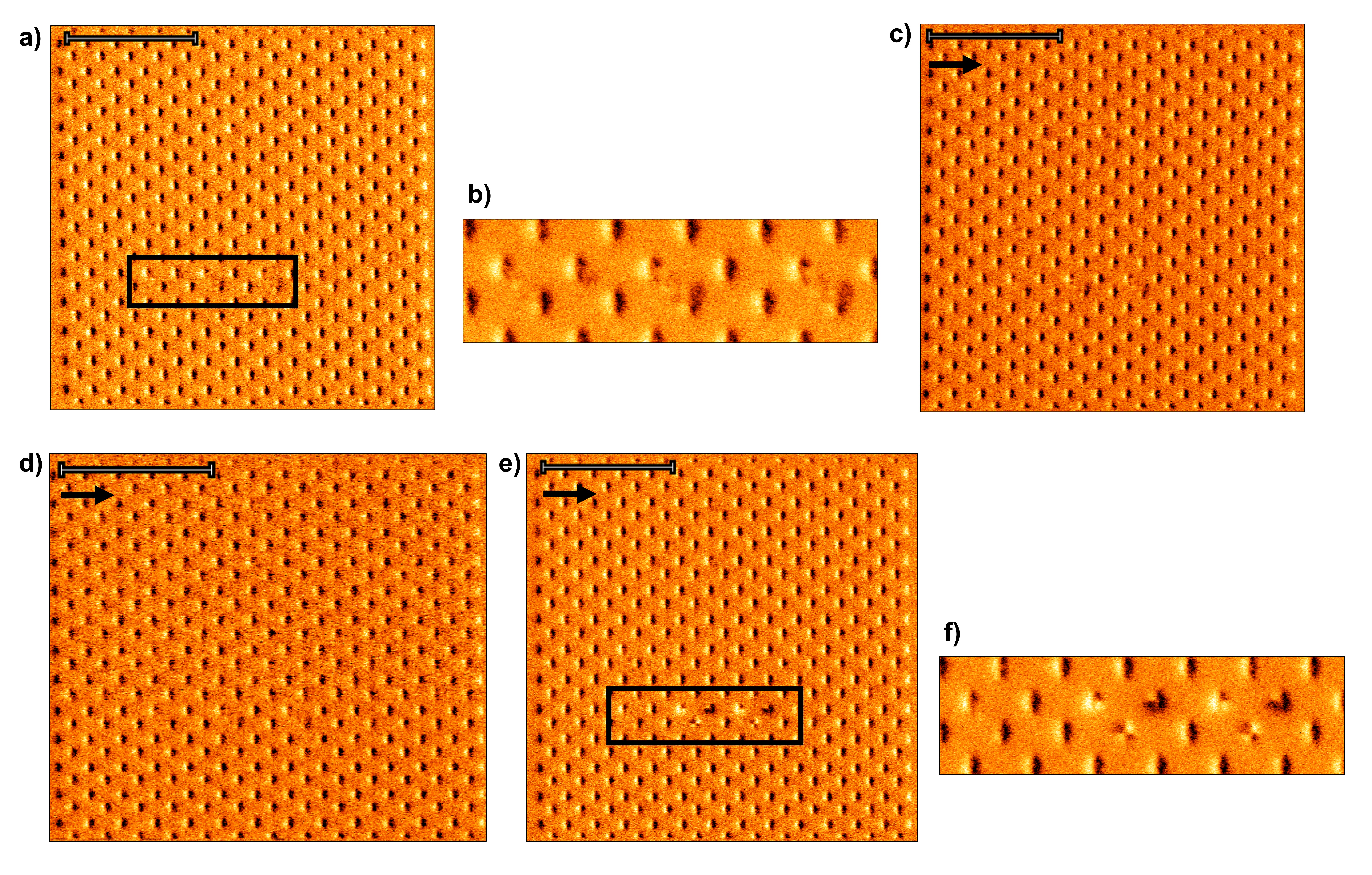}
\caption{\label{figS10} \textbf{a)} MFM image of initially -X saturated Array 2 after laser-writing has induced two double-vortices in a wide bar row, next to a possible third in a potentially partially damaged bar. The scale bar is 5 \textmu m long. \textbf{b)} High-resolution image of the section of a) highlighted by the black box. \textbf{c)} Image of the subsequent microstate of the array after application of a magnetic field of magnitude 74\% H\textsubscript{c-start} in the +X direction, as indicated by the arrow. \textbf{d)} Image of the subsequent microstate of the array after application of a magnetic field of magnitude 78\% H\textsubscript{c-start} in the +X direction, as indicated by the arrow. \textbf{e)} Image of the subsequent microstate of the array after application of a magnetic field of magnitude 82\% H\textsubscript{c-start} in the +X direction, as indicated by the arrow. \textbf{f)} High-resolution image of the section of e) highlighted by the black box.}
\end{figure*}

\begin{figure*}
    \includegraphics[width=\textwidth]{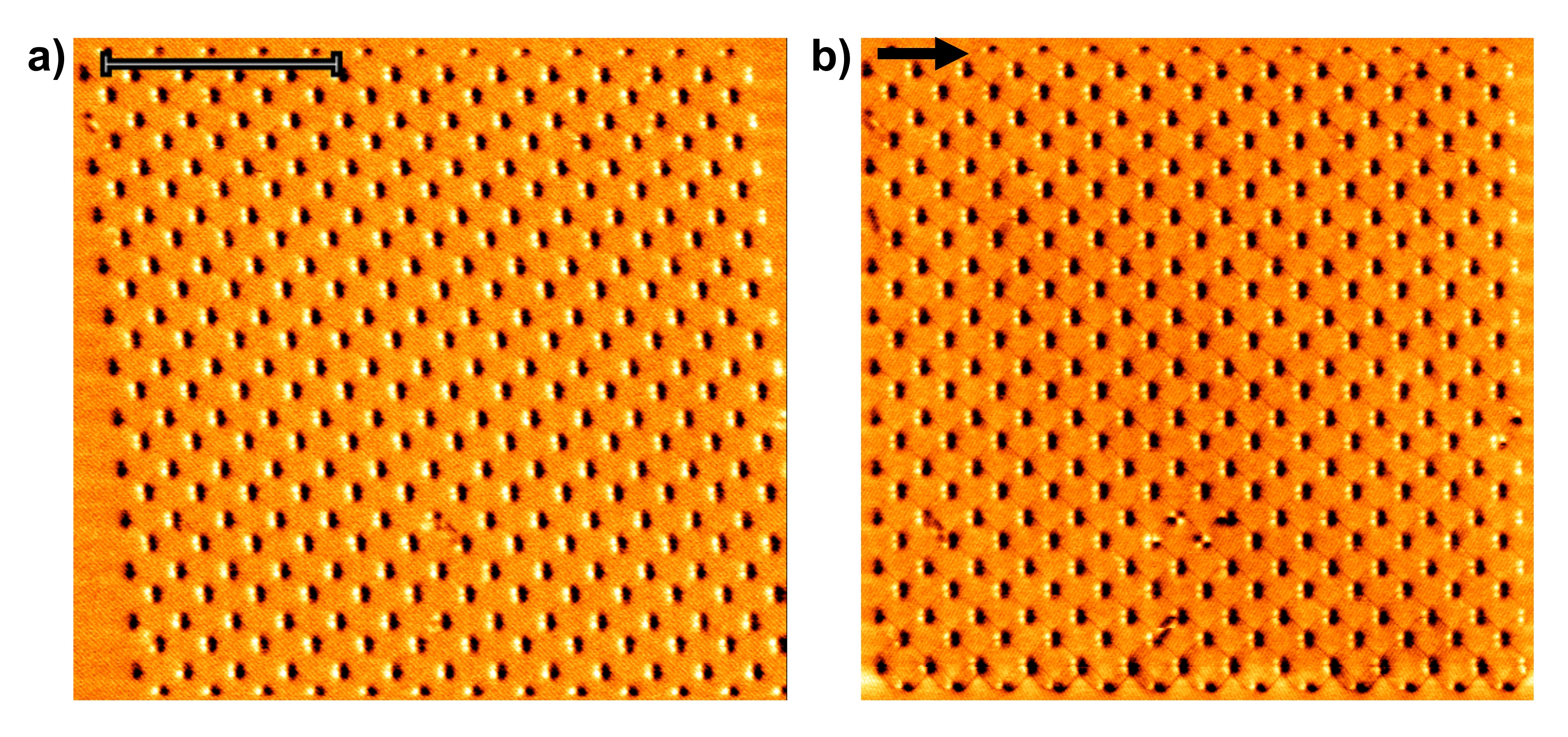}
\caption{\label{figS11} \textbf{a)} MFM image of initially -X saturated Array 1 after laser-writing has induced a single double-vortex. The scale bar is 5 \textmu m long. \textbf{b)} Image of the subsequent microstate of the array after application of a magnetic field of magnitude 91\% H\textsubscript{c-start} in the +X direction, as indicated by the arrow. This MFM image is a collage of two separate scans to display to whole array.}
\end{figure*}

\begin{figure*}
    \includegraphics[width=\textwidth]{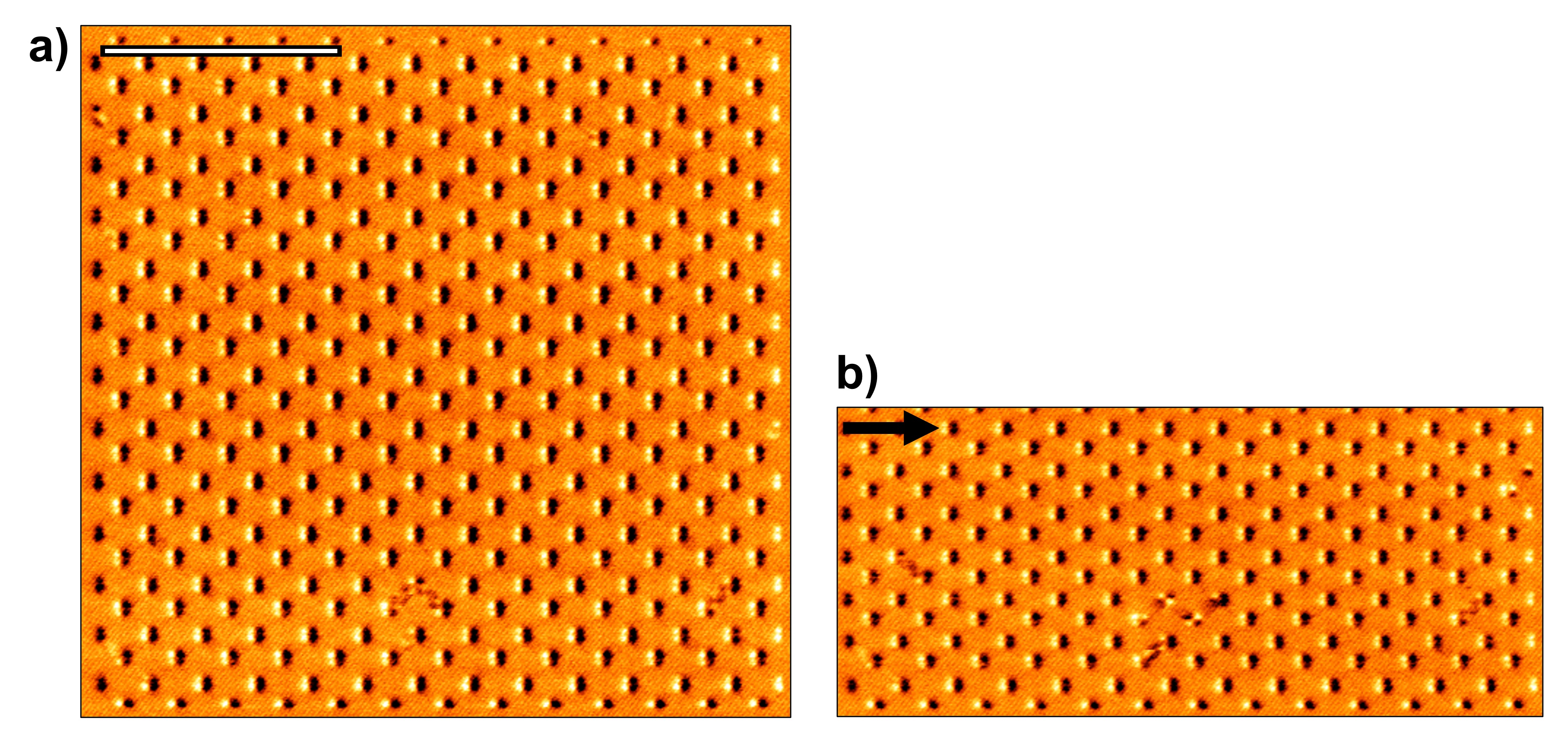}
\caption{\label{figS12} \textbf{a)} MFM image of initially -X saturated Array 1 after laser-writing has induced three double-vortices in a wide bar row. The scale bar is 5 \textmu m long. \textbf{b)} Image of the subsequent microstate of the array after application of a magnetic field of magnitude 91\% H\textsubscript{c-start} in the +X direction, as indicated by the arrow. Only the bottom section of the array was imaged.}
\end{figure*}

Figure \ref{figS9} a-b) show the full sized images of Array 1 from Figure 4. a) shows the array after initial saturation and laser-writing, with b) the subsequent microstate of the array after application of a magnetic field of magnitude 91\% H\textsubscript{c-start} in the +X direction. 

Figure \ref{figS10} a-b) shows Array 2 after initial saturation and laser-writing, c) after application of a magnetic field of magnitude 74\% H\textsubscript{c-start} in the +X direction, d) after application of a magnetic field of magnitude 78\% H\textsubscript{c-start} in the +X direction, and e-f) after application of a magnetic field of magnitude 82\% H\textsubscript{c-start} in the +X direction. Figures \ref{figS11} and \ref{figS12} show further laser-writing and field application of Array 1, with a) showing the array after initial saturation and laser-writing, and b) the application of magnetic fields of magnitude 91\% H\textsubscript{c-start} in the +X direction.

The experimental switching fields of different arrays were measured using a commercial Durham Magneto Optics nanoMOKE2 system to measure the longitudinal Magneto-Optical Kerr Effect (`MOKE') signal, with values scaled to a reference sample, whose switching fields have been well characterised by a Physical Property Measurement System (`PPMS') system. Magnetic fields were applied to the sample via the same MOKE electromagnet, such that the percentages of H\textsubscript{c-start} could be calculated from raw voltage and measured Hall probe fields. Although there was a non-negligible error on the measured magnetic field from the Hall probe (due to remanent magnetisation, and sample positioning), using the same electromagnet and calculating the applied field as a percentage of H\textsubscript{c-start}, then scaling to the reference sample, allowed for more accurate measurement.

\begin{figure*}    
\includegraphics[width=\textwidth]{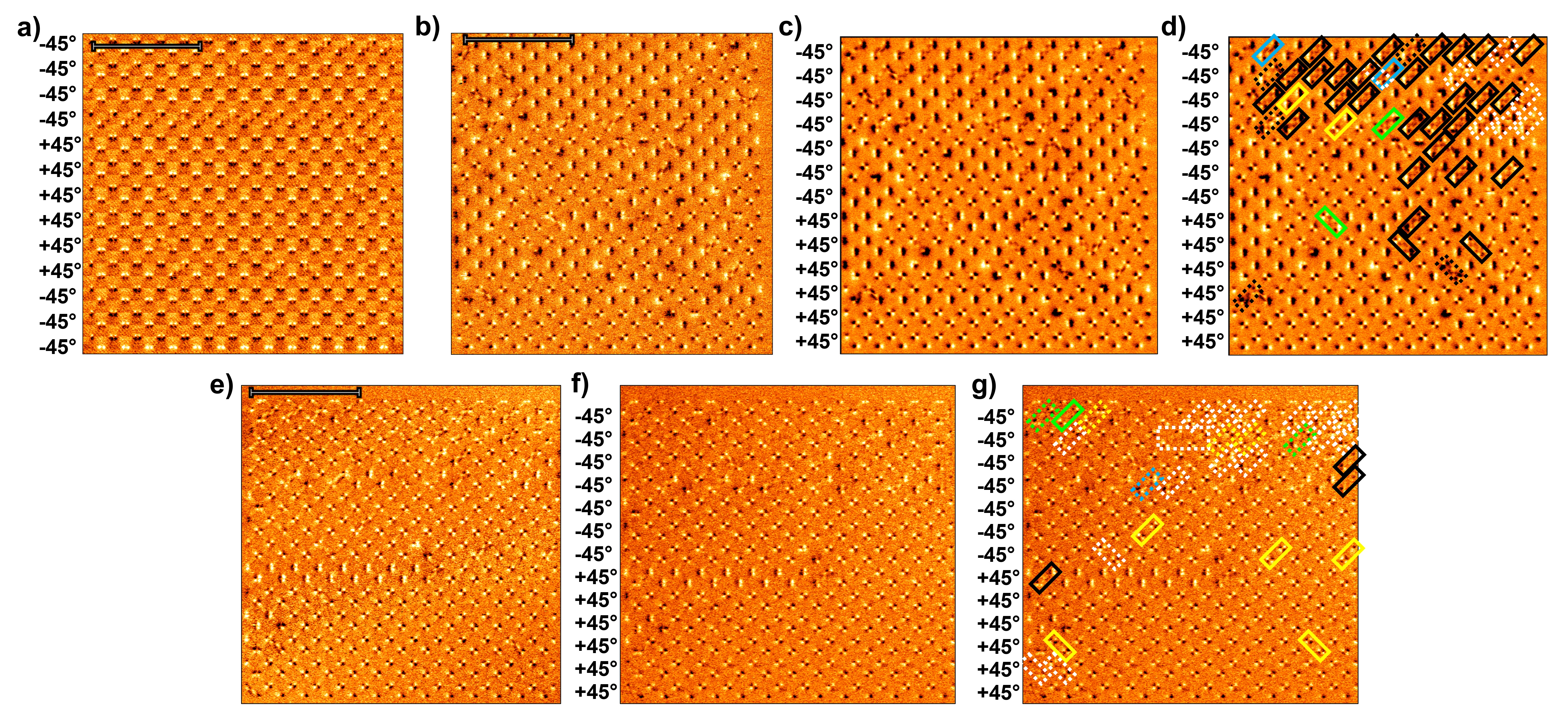}
\caption{\label{figS13}\textbf{a)} MFM image of Array 2 initially saturated in the \textbf{+y} direction and then illuminated by laser row writing. Laser polarisations are labelled. \textbf{b)} Array 2 after a degaussing protocol designed to randomise the initial magnetic state of the system. \textbf{c)} The same system after laser row illuminations, with polarisations labelled. \textbf{d)} Duplicated image of c) with changed bars labelled. Solid black boxes highlight bars that have been changed from macrospins to double-vortices. Solid blue boxes highlight bars that have been changed from macrospins to single-vortices. Solid green boxes highlight bars that have been changed from single-vortices to double-vortices. Solid yellow boxes highlight bars that have been changed from single-vortices to macrospins. White dotted boxes highlight bars that appear to have changed domain texture, but either the initial state, final state, or both are unidentified. Other dotted boxes highlight suspected changes corresponding to their respective colours. \textbf{e)} Array 2 after a new degaussing protocol designed to randomise the initial magnetic state of the system. \textbf{f)} The same system after laser row illuminations, with polarisations labelled. \textbf{g)} Duplicated image of f) with changed bars labelled as before.}
\end{figure*}

Figure \ref{figS13} a) shows Array 2 after \textbf{+y} saturation and subsequent laser-writing at 4.8 mW, with laser polarisations labelled. Figure \ref{figS13} b) shows Array 2 after a horizontal degaussing procedure designed to randomise the microstate. c) shows the subsequent state of the array after laser line writing at 4.8 mW, with polarisations labelled. d) is a duplicated image of c), with nanoislands that have changed state, or are suspected to have, highlighted. Solid black boxes highlight bars that have been changed from macrospins to double-vortices. Solid blue boxes highlight bars that have been changed from macrospins to single-vortices. Solid green boxes highlight bars that have been changed from single-vortices to double-vortices. Solid yellow boxes highlight bars that have been changed from single-vortices to macrospins. White dotted boxes highlight bars that appear to have changed domain texture, but either the initial state, final state, or both are unidentified. Other dotted boxes highlight suspected changes corresponding to their respective colours. Figure \ref{figS9} e) shows Array 2 after a new horizontal degaussing procedure designed to randomise the microstate. f) shows the subsequent state of the array after laser line writing at 4.8 mW, with polarisations labelled. g) is a duplicated image of f), with nanoislands that have changed state, or are suspected to have, highlighted as before. Using a low-moment tip to avoid unwanted tip interactions can result in less magnetically clear images, such as those in e-f), but some state changes can still clearly be observed and identified. Vortex textures, particularly double-vortices, result in less magnetic contrast, increasing the difficulty of imaging.

\subsection*{Micromagnetic simulations}

\begin{figure*}
    \includegraphics[width=\textwidth]{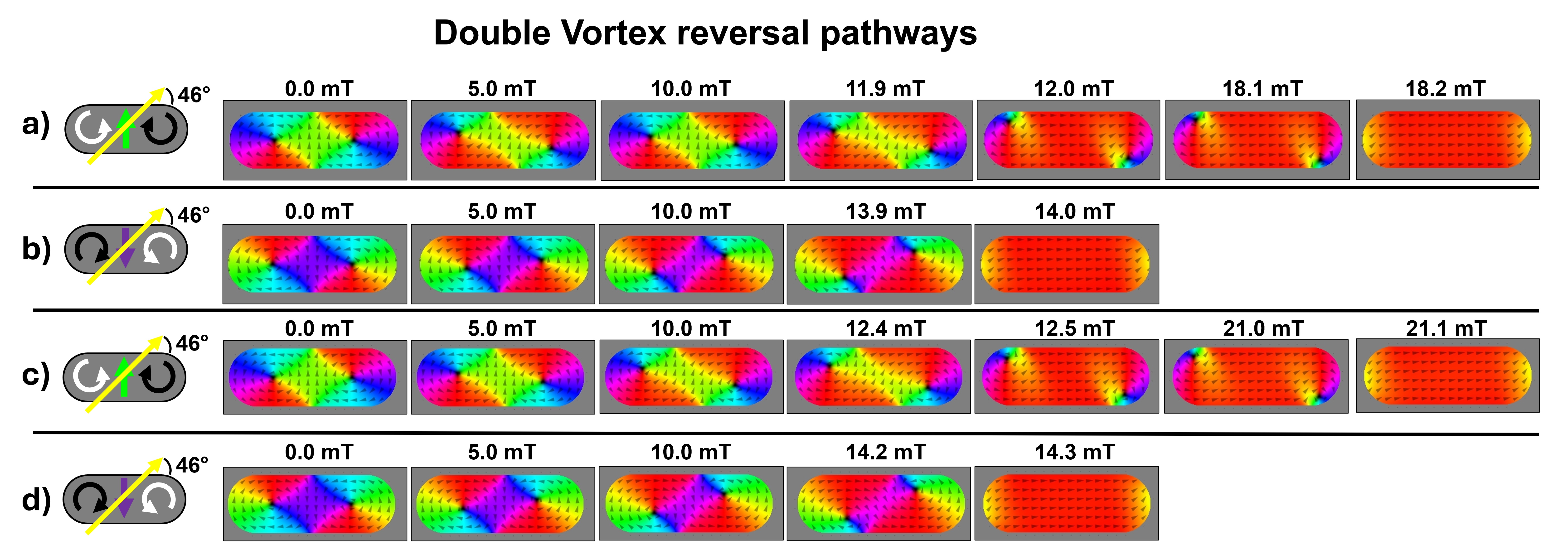}
\caption{\label{figS14}Reversal pathways for a wide bar initially in a double-vortex state from Array 1 and 2. Arrows and colours represent direction of magnetism. a) Parallel Array 1 double-vortex-to-macrospin pathway, with applied fields every 5.0 mT and at key textures labelled. b) Anti-parallel Array 1 double-vortex-to-macrospin pathway, with applied fields every 5.0 mT and at key textures labelled. c) Parallel Array 2 double-vortex-to-macrospin pathway, with applied fields every 5.0 mT and at key textures labelled. d) Anti-parallel Array 2 double-vortex-to-macrospin pathway, with applied fields every 5.0 mT and at key textures labelled.}
\end{figure*}

\begin{figure*}
    \includegraphics[width=\textwidth]{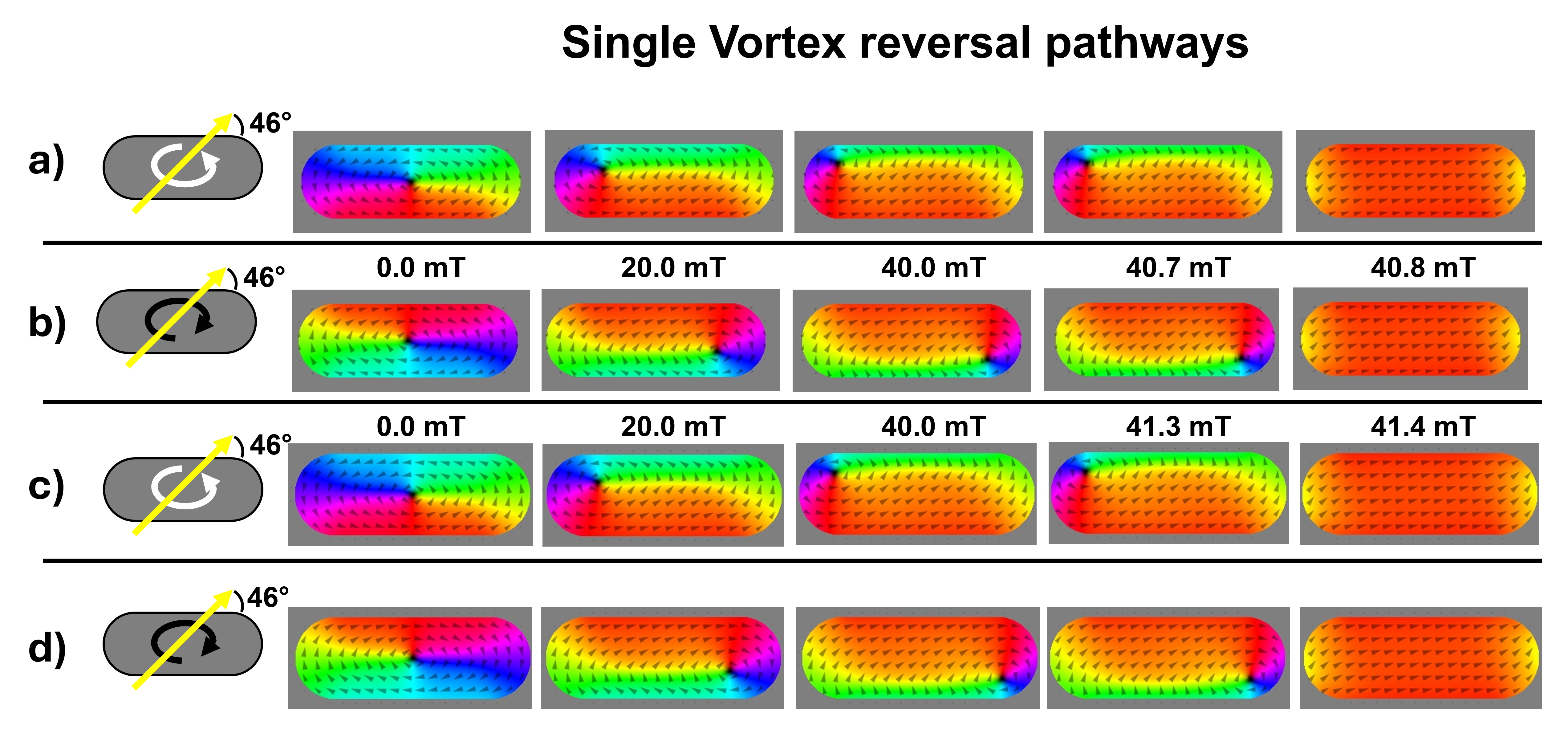}
\caption{\label{figS15}Reversal pathways for a wide bar initially in a single-vortex state from Array 1 and 2. Arrows and colours represent direction of magnetism. a) Anti-clockwise Array 1 single-vortex-to-macrospin pathway, with applied fields every 20.0 mT and at key textures labelled. b) Clockwise Array 1 single-vortex-to-macrospin pathway, with applied fields every 20.0 mT and at key textures labelled. c) Anti-clockwise Array 2 single-vortex-to-macrospin pathway, with applied fields every 20.0 mT and at key textures labelled. d) Clockwise Array 2 single-vortex-to-macrospin pathway, with applied fields every 20.0 mT and at key textures labelled.}
\end{figure*}

\begin{figure*}
    \includegraphics[width=\textwidth]{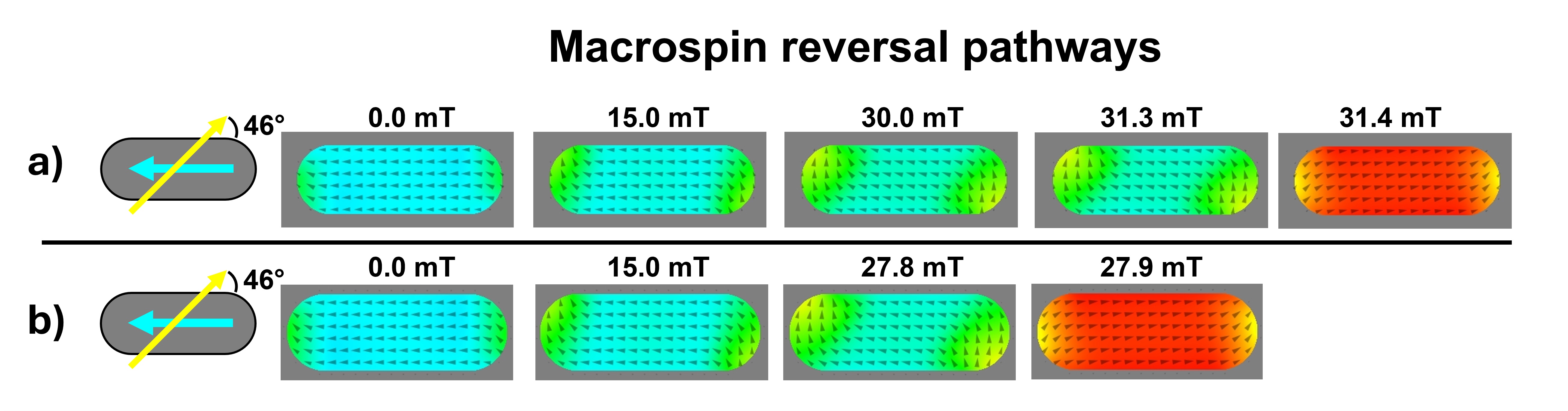}
\caption{\label{figS16}Reversal pathways for a wide bar initially in a macrospin state from Array 1 and 2. Arrows and colours represent direction of magnetism. a) Array 1 Macrospin-to-macrospin pathway, with applied fields every 15.0 mT and at key textures labelled. b) Array 2 Macrospin-to-macrospin pathway, with applied fields every 15.0 mT and at key textures labelled.}
\end{figure*}

The magnetic reversal field of isolated nanoisland bars were calculated through static micromagnetic simulations, consistent over multiple simulations, of an incrementally increasing applied magnetic field (steps of +0.1 mT) with energy minimization occurring at each field step. The discretisation of simulated cells used were 2 nm $\times$ 2 nm $\times$ 5 nm. The value of magnetic constants used and further simulation details can be found in the `Methods' section of the main text. Figures \ref{figS14} - \ref{figS16} show the evolution of reversal pathways for different magnetic textures. Fields were applied at angles of 44$^{\circ}$ and 46$^{\circ}$ relative to the long axis of the bar, and resulting fields averaged for each starting configuration.

\begin{figure*}
    \includegraphics[width=\textwidth]{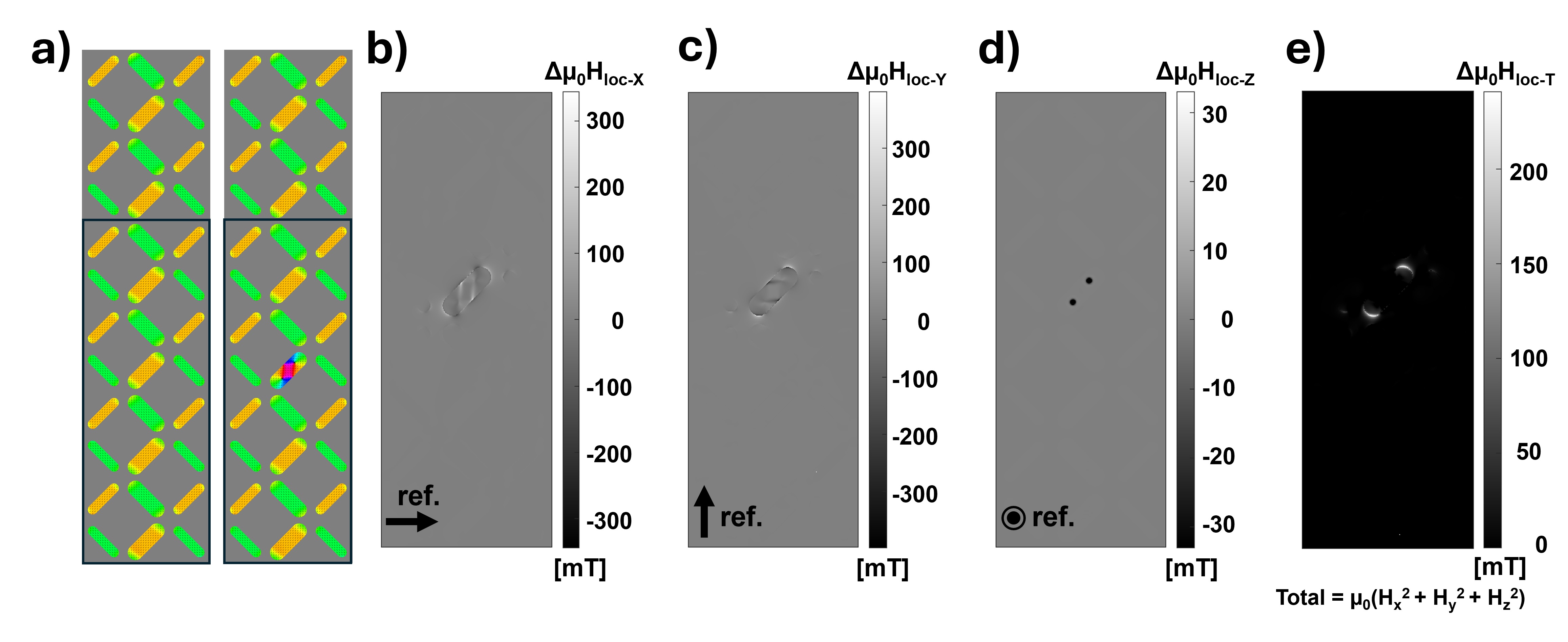}
\caption{\label{figS17} \textbf{a)} Relaxed micromagnetic states used to calculate local field reduction from neighbouring double-vortex. The left-hand image shows the system in an all-macrospin state, while the right-hand image shows the system with a double-vortex. The dimensions of the wide nanoislands correspond to those of Array 5. Multiple macrospin nanoislands were simulated within the system to prevent unwanted edge-effects. \textbf{b-d)} Plot of matching array section to that shown by the black box in a), with greyscale intensity representing the difference in demagnetising field along the reference directions indicated by the annotated arrows between the two states, where \textDelta H\textsubscript{loc} $<$ 0 represents a reduction in demagnetising field for the vortex state, in comparison to the macrospin state, with respect to the reference direction. \textbf{e)} shows the square root of the sum of each component's squared difference. Note that resulting fields are calculated in the X-Y plane of the magnets.}
\end{figure*}

\begin{figure*}
    \includegraphics[width=\textwidth]{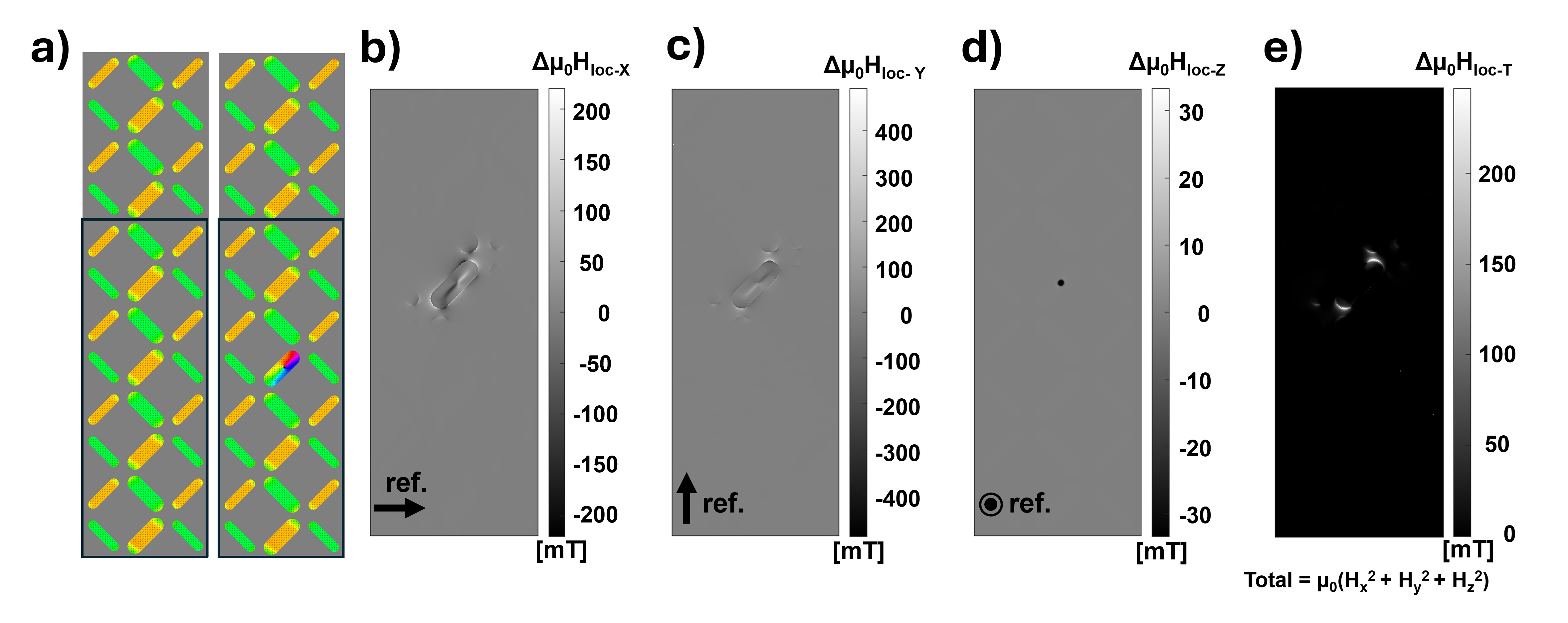}
\caption{\label{figS18} \textbf{a)} Relaxed micromagnetic states used to calculate local field reduction from neighbouring single-vortex. The left-hand image shows the system in an all-macrospin state, while the right-hand image shows the system with a single-vortex. The dimensions of the wide nanoislands correspond to those of Array 5. Multiple macrospin nanoislands were simulated within the system to prevent unwanted edge-effects. \textbf{b-d)} Plot of matching array section to that shown by the black box in a), with greyscale intensity representing the difference in demagnetising field along the reference directions indicated by the annotated arrows between the two states, where \textDelta H\textsubscript{loc} $<$ 0 represents a reduction in demagnetising field for the vortex state, in comparison to the macrospin state, with respect to the reference direction. \textbf{e)} shows the square root of the sum of each component's squared difference. Note that resulting fields are calculated in the X-Y plane of the magnets.}
\end{figure*}

Figure \ref{figS17} a) shows the simulated section of Array 5 with a double-vortex, and in an all-macrospin state. b-d) Shows the difference in demagnetising field along a reference direction between these two states for X, Y, and Z components (of the region in a) enclosed by the black box), and e) the square root of the sum of each component's squared difference, where \textDelta H\textsubscript{loc} $<$ 0 represents a reduction in demagnetising field for the vortex state, in comparison to the macrospin state. Figure \ref{figS18} likewise shows the same panels for a single-vortex vs macrospin state.

\label{Bibliography}
\bibliography{bib.bib}
\relax
\end{document}